\documentclass[12pt,preprint]{aastex}

\def\grb{GRB\thinspace{000418}}
\def\ra#1#2#3{#1$^{\rm h}$#2$^{\rm m}$#3$^{\rm s}$}
\def\dec#1#2#3{$#1^\circ#2'#3''$}

\begin{document}

\def\cit{1}
\def\ifa{2}
\def\vla{3}
\def\uwm{4}
\def\uha{5}

\title{\large A Submillimeter and Radio Survey of Gamma-Ray Burst 
Host Galaxies: A Glimpse into the Future of Star Formation 
Studies}   

\author{
E.    Berger\altaffilmark{\cit},
L. L. Cowie\altaffilmark{\ifa},
S. R. Kulkarni\altaffilmark{\cit},
D. A. Frail\altaffilmark{\vla},
H.    Aussel\altaffilmark{\ifa},
\&\ A. J. Barger\altaffilmark{\ifa,\uwm,\uha}
}

\altaffiltext{\cit}{Division of Physics, Mathematics and Astronomy,
        105-24, California Institute of Technology, Pasadena, CA
        91125}
\altaffiltext{\ifa}{Institute for Astronomy, University of Hawaii, 
	2680 Woodlawn Drive, Honolulu, HI 96822}
\altaffiltext{\vla}{National Radio Astronomy Observatory, Socorro,
        NM 87801}
\altaffiltext{\uwm}{Department of Astronomy, University of 
	Wisconsin-Madison, 475 North Charter Street, Madison, WI 
	53706}
\altaffiltext{\uha}{Department of Physics and Astronomy, University of 
	Hawaii, 2505 Correa Road, Honolulu, HI 96822}

\begin{abstract}
We present the first comprehensive search for submillimeter and 
radio emission from the host galaxies of twenty well-localized 
$\gamma$-ray bursts (GRBs).  With the exception of a single source, 
all observations
were undertaken months to years after the GRB explosions to ensure
negligible contamination from the afterglows.  We detect the host
galaxy of GRB\,000418 in both the sub-mm and radio, and the host
galaxy of GRB\,000210 only in the sub-mm.  These observations, in
conjunction with the previous detections of the host galaxies of
GRB\,980703 and GRB\,010222, indicate that about $20\%$ of GRB host
galaxies are ultra-luminous ($L>10^{12}$ L$_\odot$) and have star
formation rates of about $500$ M$_\odot$ yr$^{-1}$.  As an ensemble,
the non-detected hosts have a star formation rate of about $100$
M$_\odot$ yr$^{-1}$ ($5\sigma$) based on their radio emission.  The
detected and ensemble star formation rates exceed the
optically-determined values by an order of magnitude, indicating
significant dust obscuration.  In the same vein,
the ratio of bolometric dust luminosity to UV luminosity for the hosts
detected in the sub-mm and radio ranges from $\sim 20-800$, and
follows the known trend of increasing obscuration with increasing
bolometric luminosity.  We also show that, both as a sample and
individually, the GRB host galaxies have bluer $R-K$ colors as
compared with galaxies selected in the sub-mm in the same redshift
range.  This possibly indicates that the stellar populations in the
GRB hosts are on average younger, supporting the massive stellar
progenitor scenario for GRBs, but it is also possible that GRB hosts
are on average less dusty.  Beyond the specific results presented in
this paper, the sub-mm and radio observations serve as an
observational proof-of-concept in anticipation of the upcoming launch
of the SWIFT GRB mission and SIRTF.  These new facilities will
possibly bring GRB host galaxies into the forefront of star formation
studies. 
\end{abstract}


\section{Introduction}
\label{sec:intro} 

One of the major thrusts in modern cosmology is an accurate census of
star formation and star-forming galaxies in the Universe.  This
endeavor forms the backbone for a slew of methods (observational,
analytical, and numerical) to study the process of galaxy formation
and evolution over cosmic time.  To date, star-forming galaxies
have been selected and studied mainly in two observational windows:
the rest-frame ultraviolet (UV), and rest-frame radio and far-infrared
(FIR).  For galaxies at high redshift these bands are shifted into the
optical and radio/sub-mm, allowing observations from the ground.
Still, the problem of translating the observed emission to star
formation rate (SFR) involves large uncertainty.  This is partly
because each band traces only a minor portion of the total energy
output of stars.  Moreover, the optical/UV band is significantly
affected by dust obscuration, thus requiring order of magnitude
corrections, while the sub-mm and radio bands lack sensitivity, and
therefore uncover only the most prodigiously star-forming galaxies. 

The main result that has emerged from star formation surveys over the
past few years is exemplified in the so-called Madau diagram.  Namely,
the SFR volume density, $\rho_{\rm SFR}(z)$, rises steeply to $z\sim
1$, and seemingly peaks at $z\sim 1-2$.  There is still some debate
about the how steep the rise is, with values ranging from
$(1+z)^{1.5}$ \citep{wcb+02} to $(1+z)^{4}$ (e.g.~\citealt{mfd+96}).  
The evolution beyond $z\sim 2$ is even less clear since optical/UV
observations indicate a decline \citep{mfd+96},
while recent sub-mm observations argue for a flat $\rho_{\rm SFR}(z)$
to higher redshift, $z\sim 4$ \citep{bcr00}.  Consistency with this
trend can be obtained by invoking large dust corrections in the
optical/UV \citep{sag+99}.  For general reviews of star formation
surveys we refer the reader to \citet{ken98}, \citet{as00}, and
\citet{bsi+02}. 

Despite the significant progress in this field, our current
understanding of star formation and its redshift evolution is still
limited by the biases and shortcomings of current optical/UV, sub-mm,
and radio selection techniques.  In particular,
despite the fact that the optical/UV band is sensitive to galaxies
with modest star formation rates (down to a fraction of a M$_\odot$
yr$^{-1}$) at high redshift, these surveys may miss the most dusty,
and vigorously star-forming galaxies.
Moreover, it is not clear if the simple, locally-calibrated
prescriptions for correcting the observed {\it un-obscured} SFR for
dust extinction (e.g.~\citealt{mhc99}), hold at high redshift; even if
they do, these prescriptions involve an order of magnitude correction.
Finally, the optical/UV surveys are magnitude limited, and therefore
miss the faintest sources.  

Sub-mm surveys have uncovered a population of highly dust-extincted
galaxies, which are usually optically faint, and have star formation
rates of several hundred M$_\odot$ yr$^{-1}$ (e.g.~\citealt{sib97}).
However, unlike optical/UV surveys, sub-mm surveys are severely
sensitivity limited, and only detect galaxies with $L_{\rm bol}\gtrsim
10^{12}$ L$_\odot$.  More importantly, current sub-mm bolometer arrays
(such as SCUBA) have large beams on the sky ($\sim 15$ arcsec) making
it difficult to unambiguously identify optical counterparts (which are
usually faint to begin with), and hence measure the redshifts
\citep{sib+02}; in fact, of the $\sim 200$ sub-mm galaxies identified
to date, only a handful have a measured redshift. 
Finally, translating the observed sub-mm emission to a SFR requires
significant assumptions about the temperature of the dust, and the
dust emission spectrum (e.g.~\citealt{bsi+02}).   

Surveys at decimeter radio wavelengths also suffer from low
sensitivity, but the high astrometric accuracy afforded by synthesis  
arrays such as the VLA allows a sub-arcsec localization of the 
radio-selected galaxies.  As a result, it is easier to identify 
the optical counterparts of these sources.  Recently, this approach
has been used to pre-select sources for targeted sub-mm observations
resulting in an increase in the sub-mm detection rate
\citep{bcr00,cls+02} and redshift determination \citep{cha+03}.
However, this method is biased
toward finding luminous (high SFR) sources since it requires an
initial radio detection.  An additional problem with radio, even more
than with
sub-mm, selection is contamination by active galactic nuclei (AGN).
An examination of the X-ray properties of radio and sub-mm selected
galaxies reveals that of the order of $20\%$ can have a significant
AGN component \citep{bcm+01}.

The most significant problem with current star formation studies,
however, is that the link between the optical and sub-mm/radio samples
is still not well understood.  The Hubble Deep Field provides a clear
illustration: the brightest sub-mm source does not appear to have an
optical counterpart \citep{hsd+98}, and only recently a detection has
been claimed in the near-IR ($K\approx 23.5$ mag; \citealt{dmy+02}).
Along the same line, sub-mm observations of the optically-selected
Lyman break galaxies have resulted in very few detections
\citep{css+00,prb+00,css+02}, and even the brightest Lyman break
galaxies appear to be faint in the sub-mm band \citep{blg+01}.  In
addition, there is 
considerable diversity in the properties of optical counterparts to
sub-mm sources, ranging from galaxies which are faint in both the
optical and near-IR (NIR) to those which are bright in both bands
\citep{isb+00,sib+02}.

As a result of the unclear overlap, and the sensitivity and dust
problems in the sub-mm and optical surveys, the fractions of global
star formation in the optical and sub-mm/radio selected galaxies is
not well constrained.  It is therefore not clear if the majority of
star formation takes place in ultra-luminous galaxies with very
high star formation rates, or in the more abundant lower luminosity
galaxies with star formation rates of a few M$_\odot$ yr$^{-1}$.
Given the difficulty with redshift identification of sub-mm galaxies,
the redshift distribution of dusty star forming galaxies remains
highly uncertain. 

One way to alleviate some of these problems is to study a sample of
galaxies that is immune to the selection biases of current optical/UV
and sub-mm/radio surveys, and which may draw a more representative
sample of the underlying distribution of star-forming galaxies.  The
host galaxies of $\gamma$-ray bursts (GRBs) may provide one such
sample.   

The main advantages of the sample of GRB host galaxies are: (i) The
galaxies are selected with no regard to their emission properties in
any wavelength regime, (ii) the dust-penetrating power of the
$\gamma$-ray emission results in a sample that is completely unbiased
with respect to the global dust properties of the hosts, (iii) GRBs
can be observed to very high redshifts with existing missions
($z\gtrsim 10$; \citealt{lr00}), and as a result volume corrections
for the star formation rates inferred from their hosts are trivial,
(iv) the redshift of the galaxy can be determined via absorption
spectroscopy of the optical afterglow, or X-ray spectroscopy allowing
a redshift measurement of arbitrarily
faint galaxies (the current record-holder is the host of GRB\,990510
with $R=28.5$ mag and $z=1.619$; \citealt{vfk+01}), and (v) since
there is excellent circumstantial evidence linking GRBs to massive
stars (e.g.~\citealt{bkd02}, the sample of GRB hosts is expected to
trace global star formation \citep{bn00}.

Of course, the sample of GRB hosts is not immune from its own problems
and potential biases.  The main problem is the relatively small size
of the sample in comparison to both the optical and sub-mm
samples\footnotemark\footnotetext{Currently, the sample of GRB hosts
numbers about $30$ sources, and grows at a rate of about one per
month.  The upcoming SWIFT mission is expected to increase the rate to
one per $2-3$ days.} (although the number of GRB hosts with a known
redshift exceeds the number of sub-mm galaxies with a measured
redshift).  As a result, at the present it is not possible
to assess the SFR density that is implied by GRB hosts, or its
redshift evolution.  A bias towards sub-solar metallicity for GRB
progenitors (and hence their environments) has been discussed
\citep{mw99,mwh01}, but it appears that very massive stars
(e.g.~$M\gtrsim 35$ M$_\odot$) should produce black holes even at
solar metallicity.  The impact of metallicity on additional aspects of
GRB formation (e.g.~angular momentum, loss of hydrogen envelope) is
not clear at present.  Finally, given the observed dispersion in
metallicity within galaxies (e.g.~\citealt{ala01,ork+01}), it is
likely that even if GRBs require low metallicity progenitors, this
does not imply that the galaxy as a whole has a lower than average
metallicity. 

To date, GRB host galaxies have mainly been studied in the optical and
NIR bands.  With the exception of one source (GRB\,020124;
\citealt{bkb+02}), every GRB localized to a sub-arcsecond position has 
been associated with a star-forming galaxy \citep{bkd02}.  These galaxies
range from $R\approx 22-29$ mag, have a median redshift, $\langle
z\rangle\sim 1$, and are generally typical of star-forming galaxies at
similar redshifts in terms of morphology and luminosity
\citep{dkb+01}, with star formation rates from optical spectroscopy of
$\sim 1-10$ M$_\odot$ yr$^{-1}$.  At the same time, there are hints
for higher than average ratios of [Ne\,III]\,3869 to [O\,II]\,3727,
possibly indicating the presence of massive stars \citep{dkb+01}.
Only two host galaxies have been detected so far in the radio
(GRB\,980703; \citealt{bkf01}) and
sub-mm (GRB\,010222; \citealt{fbm+02}).  

Here we present sub-mm and radio observations of a sample of $20$ GRB
host galaxies, ranging in redshift from about $0.4$ to $4.5$
(\S\ref{sec:obs}); one of the 20 sources is detected
with high significance in both the sub-mm and radio bands, and an
additional source is detected in the sub-mm (\S\ref{sec:res}).  We
compare the detected sub-mm and radio host galaxies to
local and high-$z$ ultra-luminous galaxies in \S\ref{sec:sed}, and
derive the SFRs in \S\ref{sec:sfr}.  We then compare the inferred SFRs
of the detected host galaxies, and the ensemble of
undetected hosts, to optical estimates in \S\ref{sec:opt}.
Finally, we compare the optical properties of the GRB host
galaxies to those of sub-mm and radio selected star-forming galaxies
(\S\ref{sec:comp}).

\section{Observations}
\label{sec:obs}

\subsection{Target Selection}
\label{sec:target}

At the time we conducted our survey, the sample of GRB host galaxies
numbered 25, twenty of which had measured redshifts.  These host
galaxies were localized primarily based on optical afterglows, but
also using the radio and X-ray afterglow emission.  Of the 25 host
galaxies we observed eight in both the radio and sub-mm, seven in the
radio, and five in the sub-mm.  The galaxies were drawn from the list
of 25 hosts at random, constrained primarily by the availability of
observing time.  Thus, the sample presented here does not suffer from
any obvious selection biases, with the exception of detectable
afterglow emission in at least one band. 

Sub-mm observations of GRB afterglows, and a small number of host
galaxies have been undertaken in the past.  Starting in 1997,
\citet{stv+99} and \citet{stw+01} have searched for sub-mm emission
from the afterglow of thirteen GRBs.  While they did not detect
any afterglow emission, these authors used their observations to place
constraints on emission from eight host galaxies, with typical
$1\sigma$ rms values of 1.2 mJy.  Since these were
target-of-opportunity observations, they were not always undertaken
in favorable observing conditions.

More recently, \citet{bbt+02} reported targeted sub-mm observations of
the host galaxies of four optically-dark GRBs (i.e.~GRBs lacking an
optical afterglow).  They focused on these particular sources since
one explanation for the lack of optical emission is obscuration by
dust, which presumably points to a dusty host.  None of the hosts
were detected, with the possible exception of GRB\,000210 (see
\S\ref{sec:0210}), leading the authors to conclude that the hosts of
dark bursts are not necessarily heavily dust obscured.

Thus, the observations presented here provide the most comprehensive
and bias-free search for sub-mm emission from GRB host galaxies, and
the first comprehensive search for radio emission.

\subsection{Submillimeter Observations}
\label{sec:submm}

Observations in the sub-mm band were carried out using the
Sub-millimeter Common User Bolometer Array (SCUBA; Holland et
al. 1999) on the James Clerk Maxwell Telescope
(JCMT\footnotemark\footnotetext{The JCMT is operated by the Joint
Astronomy Centre on behalf of the Particle Physics and Astronomy
Research Council of the UK, the Netherlands Organization for
Scientific Research, and National Research Council of Canada}).  We
observed the positions of thirteen well-localized GRB afterglows with 
the long (850 $\mu$m) and short (450 $\mu$m) arrays.  The
observations, summarized in Table~\ref{tab:submm}, were conducted in
photometry mode with the standard nine-jiggle pattern using the
central bolometer in each of the two arrays to observe the source.  In
the case of GRB\,000301C we used an off-center bolometer in each array
due to high noise levels in the central bolometer.  

To account for variations in the sky brightness, we used a standard
chopping of the secondary mirror between the on-source position and a
position 60 arcsec away in azimuth, at a frequency of 7.8125 Hz.  In
addition, we used a two-position beam switch (nodding), in which the
beam is moved off-source in each exposure to measure the sky.
Measurements of the sky opacity (sky-dips) were taken approximately
every two hours, and the focus and array noise were checked at least
twice during each shift. 

The pointing was checked approximately once per hour using several
sources throughout each shift, and was generally found to vary by
$\lesssim 3$ arcsec (i.e.~less than one quarter of the beam size).
All observations were  performed in band 2 and 3 weather with
$\tau_{\rm 225\,{\rm GHz}}\approx 0.05-0.12$.

The data were initially reduced with the SCUBA Data Reduction Facility
(SURF) following the standard reduction procedure.  The off-position
pointings were subtracted from the on-position pointings to account
for chopping and nodding of the telescope.  Noisy bolometers were
removed to facilitate a more accurate sky subtraction (see below),
and the data were then flat-fielded to account for the small
differences in bolometer response.  Extinction correction was
performed using a linear interpolation between skydips taken before
and after each set of on-source scans.  

In addition to the sky subtraction offered by the nodding and
chopping, short-term sky contributions were subtracted by using all
low-noise 
off-source bolometers (sky bolometers).  This procedure takes
advantage of the fact that the sky contribution is correlated across
the array.  As a result, the flux in the sky bolometers can be used to
assess the sky contribution to the flux in the on-source bolometer.
This procedure is especially crucial when observing weak sources,
since the measured flux may be dominated by the sky.  We implemented
the sky subtraction using SURF and our own routine using MATLAB.
We found that in general the SURF sky subtraction under-estimated 
the sky contribution, and as a result over-estimated the source
fluxes.  We therefore used the results of our own analysis routine.
For this  
purpose we calculated the median value of the two (three) outer rings 
of bolometers in the 850 $\mu$m (450 $\mu$m) array, after removing 
noisy bolometers (defined as those whose standard deviation over a 
whole scan deviated by more than 5$\sigma$ from the median standard
deviation of all sky bolometers).  

Following the sky subtraction, we calculated the mean and standard
deviation of the mean (SDOM) for each source in a given observing
shift.  Noisy data were eliminated in two ways.  First, the data were
binned into 25 equal time bins, and the SDOM was calculated step-wise,
i.e.~at each step the data from an additional bin were added and the
mean and SDOM were re-calculated.  In an ideal situation where the
data quality remains approximately constant, the SDOM should
progressively decrease as more data are accumulated.  However, if the
quality of the data worsens (due to deteriorating weather conditions
for example) the SDOM will increase.  We therefore removed time bins
in which the SDOM increased.  Following this procedure, we recursively
eliminated individual noisy data points using a sigma cutoff level
based on the number of data points (Chauvenet's criterion;
\citealt{tay82}) until the mean converged on a constant value.
Typically, two or three iterations were required, with only a few data
points rejected each time.  Typically, only a few percent of the data
were rejected by the two procedures.

Finally, flux conversion factors (FCFs) were applied to the resulting  
voltage measurements to convert the signal to Jy.  Using photometry
observations of Mars and Uranus, and/or secondary calibrators
(OH\,231.8+4.2, IRC+10216, and CRL\,618), we found the FCF to vary
between $180-205$ Jy/V at 850 $\mu$m, consistent with the typical
value of $197\pm 13$.  At 450 $\mu$m, the FCFs varied between
$250-450$ Jy/V.

\subsection{Radio Observations}
\label{sec:radio}

{\it Very Large Array (VLA\footnotemark\footnotetext{The VLA is
operated by the National Radio Astronomy Observatory, a facility of
the National Science Foundation operated under cooperative agreement
by Associated Universities, Inc.}):}  We observed $12$ GRB afterglow
positions with the VLA from April 2001 to February 2002.  All sources
were observed at 8.46 GHz in the standard continuum mode with $2\times
50$ MHz bands.  In addition, \grb{} was observed at 1.43 and 4.86 GHz,
and GRB\,0010222 was observed at 4.86 GHz.  In Table~\ref{tab:rad} we
provide a summary of the observations.  

In principle, since the median spectrum of faint radio sources between
1.4 and 8.5 GHz is $F_\nu\propto \nu^{-0.6}$ \citep{fkp+02}, the ideal
VLA frequency for our observations (taking into account the
sensitivity at each frequency) is 1.43 GHz.  However, we chose to
observe primarily at 8.46 GHz for the following reason.  The majority
of our observations were taken in the BnC, C, CnD, and D
configurations, in which the typical synthesized beam size is $\sim
10-40$ arcsec at 1.43 GHz, compared to $\sim 2-8$ arcsec at 8.46 GHz.
The large synthesized beam at 1.43 GHz, combined with the larger field
of view and higher intrinsic brightness of radio sources at this
frequency, would result in a significant decrease in sensitivity due to
source confusion.  Thus, we were forced to observe at higher
frequencies, in which the reduced confusion noise more than
compensates for the typical steep spectrum.  We chose 8.46 GHz rather
than 4.86 GHz since the combination of $20\%$ higher sensitivity and
$60\%$ lower confusion noise, provide a more significant impact than
the typical $30\%$ decrease in intrinsic brightness.  The 1.43 GHz
observations of \grb{} were undertaken in the A configuration, where
confusion does not play a limiting role.

For flux calibration we used the extragalactic sources 3C\,48
(J0137+331), 3C\,147 (J0542+498), and 3C\,286 (J1331+305), while the
phases were monitored using calibrator sources within $\sim 5^\circ$ 
of the survey sources. 

We used the Astronomical Image Processing System (AIPS) for data
reduction and analysis.  For each source we co-added all the
observations prior to producing an image, to increase the final
signal-to-noise.

{\it Australia Telescope Compact Array
(ATCA\footnotemark\footnotetext{The Australia Telescope is funded by
the Commonwealth of Australia for operations as a National Facility
managed by CSIRO.}):} We observed the positions of four GRB afterglows
during April 2002, in the 6A configuration at $1344$ and $1432$ MHz.
Using the 6-km baseline resulted in a significant decrease in
confusion noise, thus allowing observations at the most advantageous
frequencies.  The observations are summarized in Table~\ref{tab:rad}.

We used J1934$-$638 for flux calibration, while the phase was
monitored using calibrator sources within $\sim 5^\circ$ of the survey
sources.  The data were reduced and analyzed using the Multichannel
Image Reconstruction, Image Analysis and Display (MIRIAD) package, and
AIPS.

\section{Results}
\label{sec:res}

The flux measurements at the position of each GRB are given in
Tables~\ref{tab:submm}~and~\ref{tab:rad}, and are plotted in
Figure~\ref{fig:fluxes}.  Of the $20$ sources, only \grb{} was
detected in both the radio and sub-mm with $S/N\!>\!3$
(\S\ref{sec:0418}).  One additional source, GRB\,000210, is detected
with $S/N\!>\!3$ when combining our observations with those of
\citet{bbt+02}.  Two hosts have radio fluxes with $3\!<\!S/N\!<\!4$
(GRB\,000301C and GRB\,000926), but as we show below this is due in
part to emission from the afterglow.  

The typical $2\sigma$ thresholds are about $2$ mJy, $20$ $\mu$Jy, and
$70$ $\mu$Jy in the SCUBA, VLA, and ATCA observations, respectively.
In Figure~\ref{fig:fluxes} we plot all sources with $S/N\!>\!3$ as
detections, and the rest as $2\sigma$ upper limits.  In addition, for
the sources observed with the ATCA we plot both the 1.4 GHz upper
limits, and the inferred upper limits at 8.46 GHz assuming a typical
radio spectrum, $F_\nu\propto \nu^{-0.6}$ \citep{fkp+02}. 

One obvious source for the observed radio and sub-mm fluxes (other
than the putative host galaxies) is emission from the afterglows.  To
assess the possibility that our observations are contaminated by flux
from the afterglows we note that the observations have been undertaken
at least a year after the GRB explosion\footnotemark\footnotetext{The
single exception is GRB\,011211 for which SCUBA observations were
taken $18-20$ days after the burst.}.  On this timescale, the sub-mm
emission from the afterglow is expected to be much lower than the
detection threshold of our observations.  In fact, the brightest
sub-mm afterglows to date have only reached a flux of a few mJy (at
350 GHz), and typically exhibited a fading rate of $\sim t^{-1}$ after
about one day following the burst
\citep{stv+99,bsf+00,stw+01,fbm+02,yfh+02}.  Thus, on the
timescale of our observations, the expected sub-mm flux from the
afterglows is only $\sim 10$ $\mu$Jy, well below the detection
threshold. 

The radio emission from GRB afterglows is more long-lived, and hence
posses a more serious problem.  However, on the typical timescale of
the radio observations the 8.46 GHz flux is expected to be at most a
few $\mu$Jy (e.g.~\citealt{bsf+00}).  

In the following sections we discuss the individual detections in
the radio and sub-mm, and also provide an estimate for the radio 
emission from each afterglow.

\subsection{GRB\,000418}
\label{sec:0418}

A source at the position of \grb{} is detected at four of the five
observing frequencies with $S/N\!>\!3$.  The SCUBA source, 
which we designate SMM\,12252+2006, has a flux density of 
$F_\nu(350\,{\rm GHz})\approx 3.2\pm 0.9$ mJy, and $F_\nu(670\,{\rm 
GHz})\approx 41\pm 19$ mJy.  These values imply a spectral index, 
$\beta\approx 3.9^{+1.1}_{-1.3}$ ($F_\nu\propto \nu^\beta$), consistent
with a thermal dust spectrum as expected if the emission is due 
to obscured star formation.  

The radio source (VLA\,122519.26+200611.1), is located at
$\alpha$(J2000)=\ra{12}{25}{19.255},
$\delta$(J2000)=\dec{20}{06}{11.10}, with an uncertainty of 0.1 arcsec
in both coordinates.  This position is offset from the position of the
radio afterglow of GRB\,000418 (Berger et al. 2001) by
$\Delta\alpha=-0.40\pm 0.14$ arcsec and $\Delta\delta=-0.04\pm 0.17$
arcsec (Figure~\ref{fig:lmap}).  In comparison, the offset measured
from Keck and {\it Hubble Space Telescope} images is smaller,
$\Delta\alpha=-0.019\pm 0.066$ arcsec and $\Delta\delta=0.012\pm
0.058$ arcsec. 

VLA\,122519.26+200611.1 has an observed spectral slope $\beta=-0.17\pm
0.25$, flatter than the typical value for faint radio galaxies,
$\beta\approx -0.6$ \citep{fkp+02}, and similar to the value measured
for the host of GRB\,980703 ($\beta\approx -0.32$; \citealt{bkf01}).
The source appears to be slightly extended at 1.43 and 8.46 GHz, with
a size of about $1$ arcsec ($8.8$ kpc at $z=1.119$). 

The expected afterglow fluxes at 4.86 and 8.46 GHz at the time of our 
observations are about $5$ and $10$ $\mu$Jy, respectively
\citep{bdf+01}.  At 1.43 GHz the afterglow contribution is expected to
be about $10$ $\mu$Jy based on the 4.86 GHz
flux and the afterglow spectrum $F_\nu\propto \nu^{-0.65}$.  Thus,
despite the contribution from the afterglow, the radio detections of
the host galaxy are still significant at better than $3\sigma$ level.
Correcting for the afterglow contribution we find an actual spectral
slope $\beta=-0.29\pm 0.33$, consistent
with the median $\beta\approx -0.6$ for 8.46 GHz radio sources with a
similar flux \citep{fkp+02}.

As with all SCUBA detections, source confusion arising from the 
large beam ($D_{\rm FWHM}\approx 14$ arcsec at 350 GHz and $\approx 
6$ arcsec at 670 GHz) raises the possibility that SMM\,12252+2006 
is not associated with the host galaxy of \grb{}.  Fortunately, the 
detection of the radio source, which is located $0.4\pm 0.1$ arcsec
away from the position of the radio afterglow of \grb, indicates that
SMM\,12252+2006 and VLA\,122519.26+200611.1 are in fact the same source
--- the host galaxy of \grb.

Besides the positional coincidence of the VLA and SCUBA sources, we
gain further confidence of the association based on the spectral index
between the two bands, $\beta^{350}_{1.4}$.  This spectral index is 
redshift dependent as a result of the different spectral slopes in the
two regimes \citep{cy00,bcr00}.  We find
$\beta_{1.4}^{350}\approx 0.73\pm 0.10$, in good agreement with 
the \citet{cy00} value of $\beta_{1.4}^{350}=0.59\pm 0.16$ (for the
redshift of GRB\,000418, $z=1.119$). 

We also detect another source, slightly extended ($\theta\approx 1$
arcsec), approximately 1.4 arcsec East and 2.7 arcsec South of the
host of \grb{} (designated VLA\,122519.36+200608.4), with
$F_\nu(1.43\,{\rm GHz})=48\pm 15$ $\mu$Jy and
$F_\nu(8.46\,{\rm GHz})=37\pm 12$ $\mu$Jy (see Figure~\ref{fig:lmap}).   
This source appears to be
linked by a bridge of radio emission (with $S/N\approx 1.5$ at both
frequencies) to the host of \grb{}.  The physical separation between
the two sources, assuming both are at the same redshift, $z=1.119$, is
25 kpc.  There is no obvious optical counterpart to this source in
{\it Hubble Space Telescope} images down to about $R\sim 27.5$ mag.  

Based purely on radio source counts at 8.46 GHz \citep{fkp+02}, the
expected number of sources with $F_\nu(8.46\,{\rm GHz})\approx 37$
$\mu$Jy in a 3 arcsec radius circle is only about $2.7\times 10^{-4}$.
Thus, the coincidence of two such faint sources within 3 arcsec is
highly suggestive of an interacting system, rather than chance
superposition.   

Interacting radio galaxies with separations of about $20$ kpc, and
joined by a bridge of radio continuum emission have been observed
locally \citep{chs+93,chj+02}.  In addition, optical surveys
(e.g.~\citealt{ppc+02}) show that
a few percent of galaxies with an absolute $B$-band magnitude similar
to that of the host of \grb{}, have companions within about $30$ kpc.
The fraction of interacting systems is possibly much higher, $\sim
50\%$, in ultra-luminous systems (such as the host of \grb{}), both
locally \citep{sm96} and at high redshift (e.g.~\citealt{isb+00}).   

We note that with a separation of only 3 arcsec, the host of
GRB\,000418 and the companion galaxy fall within the SCUBA beam.
Thus, it is possible that SMM\,12252+2006 is in fact a superposition
of both radio sources.  This will change the value of
$\beta_{1.4}^{350}$ to about $0.46$.

\subsection{GRB\,980703}
\label{sec:0703}

The host galaxy of GRB\,980703 has been detected in deep radio  
observations at 1.43, 4.86, and 8.46 GHz \citep{bkf01}.  The  
galaxy has a flux $F_\nu(1.43\,{\rm GHz})=68.0\pm 6.6$ $\mu$Jy, and a
radio spectral slope $\beta=-0.32\pm 0.12$.  In addition, the radio
emission is unresolved with a maximum angular size of $0.27$ arcsec
($2.3$ kpc). 

Based on the \citet{cy00} value of $\beta_{1.4}^{350}\approx
0.54\pm 0.16$ (for the redshift of GRB\,980703, $z=0.966$), the
expected flux at 350 GHz is $F_\nu(350\,{\rm GHz})\approx
1.3_{-0.8}^{+1.9}$ mJy.  The observed ($2\sigma$) flux limit
$F_\nu(350\,{\rm GHz})<1.8$ mJy, is consistent with the expected
value.

\subsection{GRB\,010222}
\label{sec:0222}

GRB\,010222 has been detected in SCUBA and IRAM observations with a
persistent flux of about 3.5 mJy at 350 GHz and 1 mJy at 250 GHz
\citep{fbm+02}.  The persistent emission, as well as the steep spectral
slope, $\beta\approx 3.8$, indicated that while the detected emission
was partially due to the afterglow of GRB\,010222, it was dominated by
the host galaxy.  In fact, accounting for the expected afterglow
emission, we find that the host galaxy has a flux, $F_\nu(350\,{\rm
GHz})\approx 2.5\pm 0.4$ mJy.  

The radio flux predicted from the sub-mm emission \citep{cy00} is
$F_\nu(1.43\,{\rm GHz})\approx 55_{-20}^{+80}$ $\mu$Jy (for
$z=1.477$, the redshift of GRB\,010222), which corresponds to
$F_\nu(4.86\,{\rm GHz})\approx 15-60$ $\mu$Jy, and $F_\nu(8.46\,{\rm
GHz})\approx 10-45$ $\mu$Jy (assuming $\beta=-0.6$).  Therefore, our
measured values, $F_\nu(4.86\,{\rm GHz})=26\pm 8$ $\mu$Jy and
$F_\nu(8.46\,{\rm GHz})=17\pm 6$ $\mu$Jy are consistent with
the observed sub-mm emission.

The expected afterglow fluxes at 4.86 and 8.46 GHz are $3$ and $4$
$\mu$Jy, respectively, significantly lower than the measured values.
Thus, the observed flux mainly arises from the host.

\subsection{GRB\,000210}
\label{sec:0210}

Recently, \citet{bbt+02} measured a flux of $3.3\pm 1.5$ mJy
for GRB\,000210, in good agreement with our value of $2.8\pm 1.1$
mJy.  A weighted-average of the two measurements gives
$F_\nu(350\,{\rm GHz})=3.0\pm 0.9$ mJy, similar to the sub-mm flux
from the host galaxies of GRB\,000418 and GRB\,010222.  The radio flux
at the position of GRB\,000210 is $F_\nu(8.46\,{\rm GHz})=18\pm 9$
$\mu$Jy.  Based on a redshift of 0.846 \citep{pfg+02} and the sub-mm 
detection, the expected radio flux from this source \citep{cy00} is
$F_\nu(8.46\,{\rm GHz})\approx 10-50$ $\mu$Jy, consistent with the
measured value.  The expected flux of the afterglow at the time of the
radio observations is less than 1 $\mu$Jy \citep{pfg+02}.

\subsection{GRB\,980329}
\label{sec:0329}

Following the localization of GRB\,980329, \citet{stv+99}
observed the afterglow position with SCUBA and claimed the detection 
of a source with a 350 GHz flux of about $5.0\pm 1.5$ mJy on 1998,
Apr.~5.2. Subsequent observations indicated a fading trend, with a
decline to $4.0\pm 1.2$ mJy on Apr.~6.2, and $<1.8$ mJy ($2\sigma$) on 
Apr.~11.2.  Based on a comparison to the
radio flux of the afterglow, \citet{stv+99} concluded that the
detected sub-mm flux was in excess of the emission from the afterglow
itself, and therefore requires an additional component, most likely a
host galaxy.

Recently, \citet{yfh+02} re-analyzed the SCUBA data and showed
that the initial sub-mm flux was in fact only about $2.5$ mJy, and
perfectly consistent with the afterglow flux.  As a
result, it is not clear that an additional persistent component is
required.  We also re-analyzed the data from Apr.~1998 using the
method described in \S\ref{sec:submm}.  We find the following fluxes:
$2.4\pm 1.0$ mJy (Apr.~5), $2.4\pm 1.1$ mJy (Apr.~6), $1.2\pm 0.8$ mJy
(Apr.~7), $1.4\pm 0.9$ mJy (Apr.~8), and $1.6\pm 0.8$ mJy (Apr.~11).
A comparison to the results in \citet{stv+99} reveals that, with
the exception of the last epoch, they over-estimate the fluxes by
about $0.5-2.5$ mJy.  

Our observations of GRB\,980329 from September and October of 2001 
reveal a
flux, $F_\nu(350\,{\rm GHz})=1.8\pm 0.8$ mJy, indicating that the
flattening to a value of about $1.5$ mJy in the late epochs of the
Apr.~1998 observations may indicate emission from the host galaxy.

The radio observations are similarly inconclusive, with
$F_\nu(8.46\,{\rm GHz})=18\pm 8$ $\mu$Jy.  We estimate that the flux
of the afterglow at 8.46 GHz at the time of our observations is only
$1-2$ $\mu$Jy \citep{yfh+02}.

Since the redshift of GRB\,980329 is not known, we cannot assess the
expected ratio of the radio and sub-mm fluxes.

\subsection{GRB\,000926}
\label{sec:0926}

This source is detected in the VLA observations with a flux of
$F_\nu(8.46\,{\rm GHz})=33\pm 9$ $\mu$Jy ($3.7\sigma$).  The expected
flux from the afterglow at the time of the observations, $\approx 420$
days after the burst, is $10$ $\mu$Jy \citep{hys+02}.  Thus,
the observed emission exceeds the afterglow flux by $2.6\sigma$.  In
the calculations below we use a host flux of $23\pm 9$ $\mu$Jy.

\subsection{GRB\,000301C}
\label{sec:0301C}

The VLA observations of this GRB position reveal a source with
$F_\nu(8.46\,{\rm GHz})=23\pm 7$ $\mu$Jy ($3.1\sigma$).  The flux of
the afterglow at the time of the observations is about $5$ $\mu$Jy
(Berger et al. 2000).  Thus, the excess emission is significant at the
$2.5\sigma$ level.

The sub-mm emission predicted based on the \citet{cy00} relation
is $F_\nu(350\,{\rm GHz})=1.5_{-1.1}^{+3.7}$ mJy (for
$z=2.034$, the redshift of GRB\,000301C).  This value is in agreement
with the measured flux of $-1\pm 1.3$ mJy.

\section{Spectral Energy Distributions}
\label{sec:sed}

In Figure~\ref{fig:sed} we plot the radio-to-UV spectral energy
distributions (SEDs) of the detected host galaxies of GRB\,980703, 
GRB\,000418, and GRB\,010222, as well as that of Arp\,220, a
proto-typical local ultra-luminous IR galaxy (ULIRG;
\citealt{snh+84}), and ERO J164502+4626.4 (HuR\,10), a high-$z$ analog
of Arp\,220 \citep{hr94,efc+02}.  The luminosities are plotted as a
function of rest-frame frequencies, to facilitate a direct comparison.

The detected GRB hosts are brighter than Arp\,220 ($L\approx 2\times
10^{12}$ L$_\odot$, SFR$\approx 300$ M$_\odot$ yr$^{-1}$), and are
similar in luminosity to HuR\,10 ($L\approx 7\times 10^{12}$ L$_\odot$,
SFR$\sim 10^3$ M$_\odot$ yr$^{-1}$; \citealt{dgi+99}).  As such, we
expect the host galaxies to have star formation rates of a ${\rm
few}\times 100$ M$_\odot$ yr$^{-1}$, and luminosities in excess of
$10^{12}$ L$_\odot$.   

On the other hand, the optical/NIR properties of the detected GRB
hosts are distinctly different than those of Arp\,220 and HuR\,10 (as
well as other local and high-$z$ ULIRGs).  In particular, from
Figure~\ref{fig:sed} it is clear that, while the GRB host galaxies are
similar to HuR\,10 and Arp\,220 in the radio and sub-mm bands, their
optical/NIR colors (as defined for example by $R-K$) are much bluer.
Moreover, while there is a dispersion of a factor of few in the radio
and sub-mm bands between the GRB hosts, HuR\,10, and Arp\,220, the
dispersion in the optical/NIR luminosity is about two orders of
magnitude.  This indicates that there is no simple correlation between
the optical/NIR properties of GRB hosts (and possibly other galaxies)
and their FIR and radio luminosities.  In the following sections we
expound on both points.

\section{Star Formation Rates}
\label{sec:sfr}

To evaluate the star formation rates that are implied by the sub-mm
and radio measurements, we use the following expression for the
observed flux as a function of SFR \citep{yc02}:
\begin{equation}
F_\nu(\nu_{\rm obs})=\{ 25f_{\rm nth}\nu_0^{-\beta} + 
0.71\nu_0^{-0.1} + 1.3\times 10^{-6}\nu_0^3
\frac{1-{\rm exp}[-(\nu_0/2000)^{1.35}]}
{{\rm exp}(0.00083\nu_0)-1} \}
\frac{(1+z){\rm SFR}}{d_L^2}\,\,{\rm Jy}.
\label{eqn:sfr}
\end{equation}
Here, $\nu_0=(1+z)\nu_{\rm obs}$ GHz, SFR is the star formation rate
in M$_\odot$ yr$^{-1}$, $d_L$ is the luminosity distance in Mpc, and 
$f_{\rm nth}$ is a scaling factor (of order unity) which accounts for 
the difference in the conversion between synchrotron flux and SFR in
the Milky Way and other galaxies.  The first term on the
right-hand-side accounts for the fact that non-thermal synchrotron
emission arising from supernova remnants is proportional to the SFR,
while the second term is the contribution of free-free emission from
HII regions.  These two flux terms dominate in the radio regime.  

The last term in Equation.~\ref{eqn:sfr} is the dust spectrum, which
dominates in the sub-mm and FIR regimes.  In this case, the parameters
that have been chosen to characterize the spectrum are a dust
temperature, $T_d=58$ K, and a dust emissivity, $\beta=1.35$, based on
a sample of 23 IR-selected starburst galaxies with $L_{\rm
FIR}>10^{11}$ L$_\odot$ \citep{yc02}.  We note that other authors
(e.g.~\citealt{bsi+02}) favor a lower dust temperature, $T_d\approx
40$ K, which would result in somewhat different inferred star
formation rates.

To calculate $d_L$ we use the cosmological parameters $\Omega_{\rm
m}=0.3$, $\Omega_\Lambda=0.7$, and $H_0=65$ km s$^{-1}$ Mpc$^{-1}$.
We also use the typical value $\beta\approx -0.6$ for the radio
measurements \citep{fkp+02}.  In Figure~\ref{fig:fluxes} we plot
contours of constant SFR overlaid on the sub-mm and radio flux
measurements.  Our radio observations are sensitive to galaxies with
${\rm SFR}>100$ M$_\odot$ yr$^{-1}$ at $z\sim 1$, and  ${\rm
SFR}>1000$ M$_\odot$ yr$^{-1}$ at $z\sim 3$.  The sub-mm flux, on the
other hand, is relatively constant for a given SFR, independent of
$z$.  This is due to the large positive k-correction resulting from
the steep thermal dust spectrum.  Therefore, at the typical limit of
our sub-mm observations we are sensitive to galaxies with SFR$\gtrsim
500$ M$_\odot$ yr$^{-1}$. 

For the host galaxies that are detected with $S/N\!>\!3$ in the
sub-mm and radio, as well as those detected in the past
(i.e.~GRB\,980703 and GRB\,010222) we calculate the following star
formation rates: GRB\,000418 -- ${\rm SFR}_S=690\pm 200$ M$_\odot$
yr$^{-1}$, ${\rm SFR}_R=330\pm 75$ M$_\odot$ yr$^{-1}$;
GRB\,000210 -- ${\rm SFR}_S=560\pm 170$ M$_\odot$ yr$^{-1}$;
GRB\,010222 -- ${\rm SFR}_S=610\pm 100$ M$_\odot$; GRB\,980703 --
${\rm SFR}_R=180\pm 25$ M$_\odot$ yr$^{-1}$.  Here ${\rm SFR}_S$ and
${\rm SFR}_R$ are the SFRs derived from the sub-mm and radio fluxes,
respectively.   

The detections and upper limits from this survey, combined with the
detections and upper limits discussed in the literature
\citep{bkf01,vfg+01,fbm+02} indicate that about $20\%$ of all GRBs
explode in galaxies with star formation rates of ${\rm few}\times 100$
M$_\odot$ yr$^{-1}$.  At the same time, it is clear that $\sim 80\%$
of GRB host galaxies have more modest star formation rates, ${\rm
SFR}\lesssim 100$ M$_\odot$ yr$^{-1}$.  

Despite the fact that the majority of the survey sources are not
detected, we can ask the question of whether the GRB host galaxies
exhibit a significant sub-mm and/or radio emission {\it on average}.
The weighted average emission from the non-detected sources
($S/N\!<\!3$) is $\langle F_\nu(350\,{\rm GHz})\rangle=0.37\pm 0.34$
mJy, and $\langle F_\nu(8.46\,{\rm GHz})\rangle=17.1\pm 2.7$ $\mu$Jy.
This average radio flux is possibly contaminated by flux from the
afterglows at the level of about $3$ $\mu$Jy, so we use $\langle
F_\nu(8.46\,{\rm GHz})\rangle\approx 14\pm 2.7$ $\mu$Jy
($5.2\sigma$).  Therefore, as an ensemble, the GRB host galaxies
exhibit radio emission, but no significant sub-mm emission.  Using the
median redshift, $z\approx 1$, for the non-detected sample, the
average radio flux implies an average $\langle{\rm SFR_R}\rangle\approx
100$ M$_\odot$ yr$^{-1}$, while the sub-mm $2\sigma$ upper limit on
$\langle{\rm SFR_S}\rangle$ is about $150$ M$_\odot$ yr$^{-1}$. 

The average sub-mm flux can be compared to $\langle F_\nu(350\,{\rm
GHz})\rangle=0.8\pm 0.3$ mJy for the non-detected sub-mm sources in a
sample of radio pre-selected, optically faint ($I>25$ mag) galaxies
\citep{crl+01}, $\langle F_\nu(350\,{\rm GHz})\rangle=0.4\pm 0.2$
mJy for Lyman break galaxies \citep{wef+02}, or $\langle
F_\nu(350\,{\rm GHz})\rangle\approx 0.2$ mJy for optically-selected
starbursts in the Hubble Deep Field \citep{prb+00}.  Thus, it appears
that GRB host galaxies trace a somewhat fainter population of sub-mm
galaxies compared to the radio pre-selected sample, but similar to the
Lyman break and HDF samples.  This is not surprising given that the
radio pre-selection is naturally biased in favor of luminous sources.

We can further extend this analysis by calculating the average sub-mm
and radio fluxes in several redshift bins.  Here we include both
detections and upper limits.  From the sub-mm (radio) observations we
find: $\langle F_\nu\rangle=-0.2\pm 0.4$ mJy ($\langle F_\nu\rangle=
24\pm 3$ $\mu$Jy) for $z=0-1$, $\langle F_\nu\rangle=2.3\pm 0.3$ mJy
($\langle F_\nu\rangle=16\pm 4$ $\mu$Jy) for $z=1-2$, and $\langle
F_\nu\rangle=0.5\pm 0.7$ mJy ($\langle F_\nu\rangle=18\pm 5$ $\mu$Jy)
for $z>2$.  These average fluxes are marked in Figure~\ref{fig:fluxes}.  
In the
sub-mm there is a clear increase in the average flux from $z<1$ to
$z\sim 1-2$, and a flattening or decrease beyond $z\sim 2$.  In the
radio, on the other hand, The average flux is about the same in all
three redshift bins.

The average radio fluxes translate into the following star formation
rates: for $z<1$ the inferred average SFR is $\sim 110$ M$_\odot$
yr$^{-1}$, for $1<z<2$ it is $\sim 200$ M$_\odot$ yr$^{-1}$, and 
for $z>2$ it is $\sim 700$ M$_\odot$ yr$^{-1}$ (with $>3\sigma$
significance in each bin).  The sub-mm observations on the other hand,
indicate a rise from a value of $\lesssim 160$ M$_\odot$ yr$^{-1}$ for
$z<1$ to $\sim 510$ M$_\odot$ yr$^{-1}$ for $1<z<2$, followed by a
decline to $\lesssim 320$ M$_\odot$ yr$^{-1}$ for $z>2$.

\section{Comparison to Optical Observations}
\label{sec:opt}

The typical {\it un-obscured} star formation rates inferred from
optical spectroscopy are of the order of $1-10$ M$_\odot$ yr$^{-1}$
(e.g.~\citealt{dkb+01}).  In particular, the host galaxy of
GRB\,980703 has an optical SFR of about $10$ M$_\odot$ yr$^{-1}$
\citep{dkb+98}, compared to about $180$ M$_\odot$ yr$^{-1}$ from the
radio observations.  Similarly, the host of \grb{} has an optical SFR
of about $55$ M$_\odot$ yr$^{-1}$ \citep{bbk+02}, compared to about
$300-700$ 
M$_\odot$ yr$^{-1}$ based on the radio and sub-mm detections, while
the host of GRB\,000210 has an optical SFR of $\sim 3$ M$_\odot$
yr$^{-1}$ compared to about $550$ M$_\odot$ yr$^{-1}$ from the sub-mm
observations.  Finally, the average radio SFR for the non-detected
sources, $\sim 100$ M$_\odot$ yr$^{-1}$, significantly exceeds the
average optical SFR.   

The discrepancy between the optical and radio/sub-mm star formation
rates indicates that the majority of the star formation in the GRB
host galaxies that are detected in the sub-mm and radio is obscured by
dust.  It is possible that the same is true for the sample as a whole,
but this relies on the less secure average SFR in the non-detected
hosts.  The significant dust obscuration is not surprising given that
a similar trend has been noted in high-$z$ starburst galaxies, for
which the typical dust corrections (based on the UV slope technique)
are an order of magnitude \citep{mhc99}.  In this case we find similar
correction factors.

We can also assess the level of obscuration by comparing the
UV luminosity at $1600$\AA{}, $L_{1600}$, to the bolometric
dust luminosity, $L_{\rm bol,dust}$.  The ratio of these two
quantities provides a rough measure of the obscuration, while the sum
provides a rough measure of the total star formation rate \citep{as00}.  
We estimate $L_{1600}$ using the following host magnitudes: $B\approx
23.2$ mag (GRB\,980703; \citealt{bfk+98}), $R\approx 23.5$ mag
(GRB\,0000210; Piro et al. 2002), $R\approx 23.6$ mag (GRB\,000418;
\citealt{bdf+01}), and $B\approx 26.7$ mag (GRB\,010222;
\citealt{fbm+02}), and assuming a spectrum $F_\nu\propto \nu^{-2}$.  
We calculate the following values of $L_{1600}$: $6\times 10^{10}$
L$_\odot$ (GRB\,980703), $6\times 10^{9}$ L$_\odot$ (GRB\,000210),
$5\times 10^{9}$ L$_\odot$ (GRB\,000418), and $5\times 10^{10}$
L$_\odot$ (GRB\,010222).  

We estimate $L_{\rm bol,dust}$ from the radio and sub-mm observations,
using the conversion factors of \citet{as00}.  The resulting values
are: $1.3\times 10^{12}$ L$_\odot$ (GRB\,980703), $3.3\times 10^{12}$
L$_\odot$ (GRB\,000210), $4.4\times 10^{12}$ L$_\odot$ (GRB\,000418),
and $4.1\times 10^{12}$ L$_\odot$ (GRB\,010222).  Thus, $L_{\rm
bol,dust}/L_{1600}$ evaluates to: $20$ (GRB\,980703), $510$
(GRB\,000210), $810$ (GRB\,000418), and $90$ (GRB\,010222).  These
results, as well as the sample of starbursts and ULIRGs at $z\sim 1$
taken from \citet{as00} are plotted in Figure~\ref{fig:lbol}.  We note
that the GRB hosts are within the scatter of the $z\sim 1$
sample, with a general trend of increasing value of $L_{\rm
bol,dust}/L_{1600}$ (i.e.~inceasing obscuration) with increasing
$L_{\rm bol,dust}+L_{1600}$ (i.e.~inceasing SFR).  

At the same time, the particular lines of sight to the GRBs within the
sub-mm/radio bright host galaxies do not appear to be heavily
obscured.  For example, an extinction of $A_V^{\rm host}\sim 1$ mag
has been inferred for GRB\,980703 \citep{fra+02}, $A_V^{\rm host}\sim
0.4$ mag has been found for GRB\,000418 \citep{bdf+01}, and $A_V^{\rm
host}\sim 0.1$ mag has been found for GRB\,010222.  The optically-dark
GRB\,000210 suffered more significant extinction, $A_R^{\rm
host}>1.6$ mag.  In addition, the small offset of GRB\,980703 relative
to its radio host galaxy ($0.04$ arcsec; $0.3$ kpc at the redshift of
the burst), combined with the negligible extinction, indicates that
while the burst probably exploded in a region of intense star
formation, it either managed to destroy a large amount of dust in its 
vicinity, or the dust distribution is patchy.  It is beyond the scope
of this paper to evaluate the potential of dust destruction by GRBs
(see e.g.~\citealt{wd00}), but it is clear that the GRBs that exploded
in the detected sub-mm and radio host galaxies, did not occur in the
most heavily obscured star formation sites.

\section{Comparison of the Optical Properties of GRB hosts to Radio
Pre-Selected Sub-mm Galaxies}
\label{sec:comp}

As we noted in \S\ref{sec:sed}, the optical/NIR colors of the detected
GRB host galaxies are bluer than those of Arp\,220 ($R-K\approx 4$ mag)
and HuR\,10 ($I-J\approx 5.8$ mag; \citealt{dgi+99}).  In this section
we compare the $R-K$ color of GRB hosts to the $R-K$ colors of radio 
pre-selected sub-mm galaxies \citep{cha+03,lch+03} and sub-mm selected
galaxies with a known optical counterpart and a redshift
\citep{fis+98,isl+98,fis+99}. 

In Figure~\ref{fig:color} we plot $R-K$ color versus redshift for GRB
hosts and radio pre-selected sub-mm galaxies.  Before comparing the
two populations, we note that the mean $R-K$ color and redshift for
the entire GRB sample are $2.6\pm 0.6$ mag and $1.0\pm 0.3$,
respectively, and for the hosts that are detected in the sub-mm and
radio they are $2.6\pm 0.3$ mag and $1.1\pm 0.3$, respectively.  Thus,
there is no clear correlation between the optical/NIR colors of the
GRB hosts and their sub-mm/radio luminosity.

For the sub-mm sample the mean $R-K$ color and redshift are $4.6\pm
1.0$ mag and $1.8\pm 0.7$, respectively.  To facilitate a more direct
comparison with the GRB sample we calculate the mean values for
the sub-mm sample in the same redshift range as the GRB hosts:
$\langle R-K\rangle=5.1\pm 0.9$ mag and $\langle z\rangle=1.1\pm
0.3$.  Moreover, if we examine only the host galaxies that were
detected in the radio and sub-mm with high significance we find $R-K$
colors of: 2.2 mag (GRB\,000418), 2.8 mag (GRB\,980703), 2.1 mag
(GRB\,010222), and 2.6 (GRB\,000210).  The bluest sub-mm galaxies,
on the other hand, have $R-K\approx 3.1$ mag.  In general, the GRB
hosts that are detected with $S/N\!>\!3$ in the sub-mm and radio have
a distribution of $R-K$ that is indistinguishable from the general GRB
host galaxy sample.   

The obvious difference in $R-K$ color indicates that the GRB and
radio/sub-mm selections result in a somewhat different set of
galaxies.  The red colors of the sub-mm selected galaxies are not
surprising since these sources are expected to be dust obscured.  On
the other hand, the mean color of the GRB hosts is bluer by about 2.5 
mag ($2.3\sigma$ significance) compared to sub-mm galaxies in the same
redshift range, indicating a bias towards less dust obscuration. a
more patchy dust distribution, or intrinsically bluer colors. 

It is possible that there is a bias toward less dust obscuration in
the general GRB host sample because the bursts that explode in dusty
galaxies would have obscured optical afterglows, and hence no accurate
localization.  However, this is not a likely explanation since the
GRBs which exploded in the sub-mm and radio bright hosts are not
significantly dust obscured (\S\ref{sec:opt}).  Moreover, it does
not appear that the hosts of dark GRBs are brighter in the sub-mm as
expected if the dust obscuration is global \citep{bbt+02}.  Finally,
the localization of 
afterglows in the radio and X-rays allows the selection of host
galaxies even if they are dusty.  In particular, the only two GRBs in
which significant obscuration of the optical afterglow has been
inferred (GRB\,970828: \citealt{dfk+01}; GRB\,000210: Piro et
al. 2002), have been localized thanks to accurate positions from the
radio and X-ray afterglows, and have host galaxies with $R-K$ colors
of 3.7 and 2.6 mag, not significantly redder than the general
population of GRB hosts. Therefore,
a bias against dust obscured host galaxies is not the reason for
the bluer color of the sample.

An alternative explanation is that the distribution of dust in GRB
hosts is different than in the radio pre-selected and sub-mm selected
galaxies.  This may be in terms of a spatially patchy distribution,
which will allow more of the UV light to escape, or a different
distribution of grain sizes (i.e.~a different extinction law),
possibly due to a different average metallicity.  However, in both
cases it is not clear why there should be a correlation between the
dust properties of the galaxy and the occurence of a GRB.

Finally, it is possible that GRB host galaxies are preferentially in
an earlier stage of the star formation (or starburst) process.  In
this case, a larger fraction of the shorter-lived massive stars would
still be shining, and the overall color of the galaxy would be bluer
relative to a galaxy with an older population of stars.  One way to
examine the age of the stellar population is to fit population
synthesis models to the broad-band optical/NIR spectra of the host
galaxies.  This approach has recently been used by \citet{cba02}
who find some evidence that the age of the stellar population in some
GRB host galaxies (including the host of GRB\,980703) is relatively
young, of the order of $10-50$ Myr.  

This result is also expected if GRBs arise from massive stars, as
indicated by recent observations (e.g.~\citealt{bkp+02}), since in
this case GRBs would preferentially select galaxies with younger
star formation episodes.

Regardless of the exact reason for the preferential selection of bluer
galaxies relative to the radio pre-selected sub-mm population,
two results seem clear:
(i) The GRB host galaxies detected in the sub-mm and radio are
likely drawn from a population that is generally missed in current
sub-mm surveys, and (ii) GRB host galaxies may not be a completely
bias-free sample.  

The first point is particularly interesting in light of the fact that
optical observations of these host galaxies do not identify them as
particularly exceptional in terms of SFR.  Therefore, while similar 
galaxies are not necessarily missed in optical surveys, their star
formation rates are likely under-estimated.

\section{Conclusions and Future Prospects}
\label{sec:conc}

We presented the most comprehensive SCUBA, VLA, and ATCA observations of 
GRB host galaxies to date.
The host galaxy of \grb{} is the only source detected with high
significance in both the sub-mm and radio, while the host galaxy of
GRB\,000210 is detected with $S/N\approx 3.3$ in the sub-mm when we
combine our observations with those of \citet{bbt+02}.  When taken in
conjunction with the previous detections of GRB\,980703 in the radio
\citep{bkf01} and GRB\,010222 in the sub-mm \citep{fbm+02}, these
observations point to a $\sim 20\%$ detection rate in the
radio/sub-mm.  This detection rate confirms predictions for the number
of sub-mm bright GRB hosts, with $F_\nu(350\,{\rm GHz})\sim 3$ mJy,
based on current models of the star formation history assuming a large
fraction of obscured star formation \citep{rtb02}. 

The host galaxies detected in the sub-mm and radio have star formation
rates from about $200$ to $700$ M$_\odot$ yr$^{-1}$, while
statistically the non-detected sources have an {\it average} SFR of
about $100$ M$_\odot$ yr$^{-1}$.  These star formation rates exceed
the optically-inferred values by over an order of magnitude, pointing
to significant dust obscuration within the GRB host galaxies detected
in the sub-mm and radio, and possibly the sample as a whole.

Still, the optical afterglows of the bursts that exploded in the 
sub-mm/radio bright host galaxies did not suffer significant
extinction, indicating that: (i) the GRBs did not explode in regions
where dust obscuration is significant, or (ii) the UV and X-ray
emission from the afterglow destroys a significant amount of dust in
the local vicinity of the burst. 

We have also shown that GRB host galaxies, even those detected in the 
sub-mm/radio, have bluer $R-K$ colors compared to galaxies selected in
the sub-mm or radio bands in the same redshift range.  This is not
the result of an observational bias against dusty galaxies in the GRB
host sample since the afterglows of GRBs which exploded in the
radio/sub-mm bright hosts were not significantly obscured.  More
likely, this is the result of younger stellar populations in these
galaxies, or possibly a patchy dust distribution.  If the reason is
younger stellar population then this provides additional
circumstantial evidence in favor of massive (and hence
short-lived) stars as the likely progenitors of GRBs.

A potential bias of the GRB host galaxy sample is that the popular
``collapsar'' model of GRBs calls for high mass, low metallicity
stellar progenitors \citep{mw99}.  This may result in preferential
selection of low metallicity (and hence less dusty) host galaxies.
However, it appears that GRB progenitors can even have solar
metallicity, and that a very low metallicity is unfavored by the
required initial conditions for a GRB explosion.  Moreover, studies of
the Milky Way (see \citealt{sta02} for a recent review), local
galaxies (e.g~\citealt{ala01}), and high-$z$ galaxies
(e.g.~\citealt{ork+01}), indicate that there are considerable
variations in metallicity within
galaxies.  This may be especially true if several independent episodes
of star formation have occured within the galaxy.  Thus, even if there
is a bias towards low metallicity for GRB progenitors (and hence their
immediate environments) it is not obvious that this introduces a bias
in the host galaxy sample.

Nonetheless, while the observations presented in this paper clearly
indicate the potential of GRB selection of high-$z$ galaxies for the
study of star formation, a much larger sample is required to
complement existing optical and sub-mm surveys.  This may become
possible in the near future with the upcoming launch (Sep.~2003) of
SWIFT.  With an anticipated rapid ($\sim 1$ minute) and accurate
localization of about 150 bursts per year, the GRB-selected sample
will probably increase to several hundred galaxies over the next few
years.  The rapid localization would most likely result in a large
fraction of redshift measurements thanks to the bright optical
afterglows.  

In addition to the localization of a large number of GRB hosts, the
study of these galaxies (as well as those in other samples) would
greatly benefit from the advent of new facilities, such as SIRTF,
ALMA, EVLA, and the SKA.  In Figure~\ref{fig:sirtf} we again plot
the rest-frame SEDs of Arp\,220 and the sub-mm/radio bright GRB
hosts.  Overplotted on these SEDs are the $1\sigma$ sensitivities of
SIRTF, ALMA, and the EVLA for 200-sec exposures at redshifts 1 and
3, as well as the sensitivities of current instruments (VLA and
SCUBA).  

The contributions of these new facilities to star formation studies
are threefold: (i) increased sensitivity, (ii) increased resolution,
and (iii) increased frequency coverage.  These improvements will serve
to ameliorate the main limitations of present radio, sub-mm, and IR
observations (\S\ref{sec:intro}), by allowing the detection of more
representative star forming galaxies at high redshift, in addition to
a better constraint on the total dust bolometric luminosity and
accurate localizations, which would facilitate follow-ups at optical
wavelengths.  In conjunction with increasingly larger samples of
galaxies selected in the optical, the radio/sub-mm/IR, and by GRBs,
the future of star formation studies is poised for great advances and
new discoveries. 

\acknowledgments

We thank K.~Adelberger, A.~Blain, and A.~Shapley for helpful
discussions, and G.~Moriarty-Schieven for help with the data
reduction.  We also thank S.~Chapman for providing us with the
optical/NIR colors and redshifts of radio pre-selected sub-mm galaxies
prior to publication.


\clearpage
\begin{deluxetable}{lclrrr}
\tablecaption{Submillimeter Observations \label{tab:submm}}
\tablehead {
\colhead {Source}       	   &
\colhead {$z$} 			   &
\colhead {Obs.~Date}		   &
\colhead {$F_\nu(350\,{\rm GHz})$} &
\colhead {$F_\nu(670\,{\rm GHz})$} &
\colhead {$\langle F_\nu(350\,{\rm GHz})\rangle$} \\ 
\colhead {} 		&
\colhead {}		&
\colhead {(UT)}		&
\colhead {(mJy)}	&
\colhead {(mJy)}	&
\colhead {(mJy)}	
}
\startdata
GRB\,970228 & 0.695 & Nov.~1, 2001 & $-1.58\pm 1.34$ & $-21.4\pm 18.6$
& \\
	    &       & Nov.~2, 2001 & $0.42\pm 1.61$  & $-10.9\pm 21.4$
& $-0.76\pm 1.03$ \\
GRB\,970508 & 0.835 & Sep.~9, 2001 & $-1.70\pm 1.56$ & $-12.2\pm 48.4$
& \\
            &       & Sep.~10, 2001 & $-0.53\pm 1.60$ & $3.2\pm 64.8$
& \\ 
            &       & Sep.~12, 2001 & $-3.64\pm 2.43$ & $6.0\pm 34.2$
& $-1.57\pm 1.01$ \\
GRB\,971214 & 3.418 & Nov.~2, 2001 & $0.49\pm 1.11$ & $-14.2\pm 12.6$
& $0.49\pm 1.11$ \\ 
GRB\,980329 & \nodata & Sep.~13, 2001 & $1.22\pm 1.62$ & $8.6\pm 10.2$
& \\
            &         & Oct.~29, 2001 & $2.06\pm 0.99$ & $-27.4\pm 21.6$ 
& $1.83\pm 0.84$ \\
GRB\,980613 & 1.096 & Nov.~1, 2001 & $2.84\pm 1.87$ & $92.6\pm 95.9$
& \\
            &       & Nov.~2, 2001 & $2.21\pm 1.77$ & $30.3\pm 64.4$
& \\ 
            &       & Dec.~7, 2001 & $0.93\pm 1.33$ & $22.6\pm 17.6$
& $1.75\pm 0.92$ \\
GRB\,980703 & 0.966 & Sep.~10, 2001 & $-2.40\pm 1.30$ & $-22.6\pm 18.6$ 
& \\ 
            &       & Sep.~12, 2001 & $-0.84\pm 1.33$ & $-13.9\pm 10.7$ 
& $-1.64\pm 0.93$ \\
GRB\,991208 & 0.706 & Dec.~6, 2001 & $-2.65\pm 1.83$ & $9.1\pm 26.9$
& \\ 
            &       & Dec.~7, 2001 & $-0.08\pm 1.42$ & $26.0\pm 17.2$
& $-1.04\pm 1.12$ \\ 
GRB\,991216 & 1.020 & Oct.~31, 2001 & $0.09\pm 1.20$ & $-6.5\pm 21.3$
& \\ 
            &       & Nov.~3, 2001 & $1.23\pm 1.85$ & $-30.2\pm 31.1$
& \\ 
            &       & Nov.~4, 2001 & $0.73\pm 2.60$ & $25.6\pm 128.5$
& $0.47\pm 0.94$ \\ 
GRB\,000210 & 0.846 & Sep.~12, 2001 & $3.96\pm 2.27$ & $98.1\pm 48.2$
& \\
            &       & Sep.~13, 2001 & $4.34\pm 1.63$ & $70.1\pm 45.1$
& \\ 
            &       & Sep.~14, 2001 & $-0.01\pm 1.87$ & $-6.4\pm 87.1$
& $2.97\pm 0.88$ \\
GRB\,000301C & 2.034 & Dec.~29, 2001 & $1.02\pm 1.99$ & $21.4\pm 10.7$
& \\
             &       & Dec.~30, 2001 & $-2.71\pm 1.79$ & $-18.7\pm 25.1$ 
& $-1.04\pm 1.33$ \\
GRB\,000418 & 1.119 & Oct.~30, 2001 & $3.80\pm 2.11$ & $9.4\pm 56.7$
& \\
            &       & Oct.~31, 2001 & $3.59\pm 1.35$ & $65.1\pm 31.4$
& \\ 
            &       & Nov.~1, 2001 & $2.32\pm 1.46$ & $31.9\pm 26.1$
& $3.15\pm 0.90$ \\
GRB\,000911 & 1.058 & Sep.~13, 2001 & $0.56\pm 1.69$ & $4.7\pm 22.7$
& \\
            &       & Sep.~14, 2001 & $-0.37\pm 2.68$ & $-11.1\pm 41.2$ 
& \\ 
            &       & Oct.~31, 2001 & $0.95\pm 2.25$ & $-35.0\pm 66.2$ 
& \\ 
            &       & Nov.~3, 2001 & $6.73\pm 2.08$ & $56.5\pm 52.3$ 
& \\ 
            &       & Nov.~4, 2001 & $3.07\pm 1.82$ & $-49.0\pm 51.3$ 
& $2.31\pm 0.91$ \\ 
GRB\,011211 & 2.140 & Dec.~29, 2001 & $1.64\pm 1.61$ & $8.1\pm 15.2$
& \\ 
            &       & Dec.~30, 2001 & $-0.11\pm 1.60$ & $-14.3\pm 42.7$ 
& \\
            &       & Dec.~31, 2001 & $3.88\pm 2.26$ & $17.7\pm 68.0$ 
& $1.39\pm 1.01$
\enddata
\tablecomments{The columns are (left to right), (1) Source name, (2)
source redshift, (3) UT date for each observation, (4) flux density at
350 GHz, (5) flux density at 670 GHz, and (6) weighted-average flux
density at 350 GHz.}
\end{deluxetable}

\clearpage
\begin{deluxetable}{lcclcr}
\tabcolsep0.1in\footnotesize
\tablecaption{Radio Observations \label{tab:rad}}
\tablehead {
\colhead {Source}       	&
\colhead {$z$} 			&
\colhead {Telescope}		&	
\colhead {Obs.~Dates}		&
\colhead {Obs.~Freq.} 		&
\colhead {$F_\nu$} \\
\colhead {} 		&
\colhead {}		&
\colhead {}		&
\colhead {(UT)}		&
\colhead {(GHz)}	&
\colhead {($\mu$Jy)}
}
\startdata
GRB\,970828 & 0.958 & VLA & Jun.~4--7, 2001 & 8.46 & $12\pm 9$ 
\\
GRB\,980329 & \nodata & VLA & Jul.~22 -- Sep.~10, 2001 & 8.46 & $18\pm 8$ 
\\ 
GRB\,980613 & 1.096 & VLA & May 18--26, 2001 & 8.46 & $11\pm 12$
\\
GRB\,981226 & \nodata & VLA & Jul.~24 -- Oct.~15, 2001 & 8.46 & $21\pm 12$
\\
GRB\,991208 & 0.706 & VLA & Apr.~14 -- Jul.~20, 2001 & 8.46 & $21\pm 9$
\\
GRB\,991216 & 1.020 & VLA & Jun.~8 -- Jul.~13, 2001 & 8.46 & $11\pm 9$
\\
GRB\,000210 & 0.846 & VLA & Sep.~16 -- Oct.~12, 2001 & 8.46 & $18\pm 9$ 
\\
GRB\,000301C & 2.034 & VLA & Jun.~15 -- Jul.~22, 2001 & 8.46 & $23\pm 7$
\\
GRB\,000418 & 1.119 & VLA & Jan.~14 -- Feb.~27, 2002 & 1.43 & $69\pm 15$ 
\\
            &       & VLA & Dec.~8, 2001 -- Jan.~10, 2002 & 4.86 &
$46\pm 13$ \\ 
            &       & VLA & May 28 -- Jun.~3, 2001 & 8.46 & $51\pm 12$
\\
GRB\,000911 & 1.058 & VLA & Mar.~21 -- Apr.~2, 2001 & 8.46 & $6\pm 17$
\\
GRB\,000926 & 2.037 & VLA & Jun.~11 -- Jul.~12, 2001 & 8.46 & $33\pm 9$ 
\\
GRB\,010222  & 1.477 & VLA & Sep.~29 -- Oct.~13, 2001 & 4.86 & $19\pm 10$ \\
            &       & VLA & Jun.~24 -- Aug.~27, 2001 & 8.46 & $17\pm6$
\\\hline 
GRB\,990510 & 1.619 & ATCA & Apr.~28, 2002     & 1.39 & $9\pm 35$ \\
GRB\,990705 & 0.840 & ATCA & Apr.~21--22, 2002 & 1.39 & $40\pm 34$ \\
GRB\,000131 & 4.5   & ATCA & Apr.~28, 2002     & 1.39 & $52\pm 32$ \\
GRB\,000210 & 0.846 & ATCA & Apr.~27--28, 2002 & 1.39 & $80\pm 52$ 
\enddata
\tablecomments{The columns are (left to right), (1) Source name, (2)
source redshift, (3) Telescope, (4) range of UT dates for each
observation, (5) observing frequency, and (6) peak flux density at the
position of each source.}
\end{deluxetable}

\clearpage
\begin{deluxetable}{lccc}
\tabcolsep0.1in\footnotesize
\tablecaption{Derived Star Formation Rates \label{tab:sfr}}
\tablehead {
\colhead {Source}       	 &
\colhead {Submm SFR} 		 &
\colhead {Radio SFR}		 &
\colhead {Optical SFR}		 \\
\colhead {} 			 &
\colhead {(M$_\odot$ yr$^{-1}$)} &
\colhead {(M$_\odot$ yr$^{-1}$)} &
\colhead {(M$_\odot$ yr$^{-1}$)}
}
\startdata
GRB\,970228   	  & $<335$ 	 & \nodata 	& 1 \\
GRB\,970508  	  & $<380$ 	 & \nodata 	& 1 \\
GRB\,970828  	  & \nodata	 & $80\pm 60$	& 1.2 \\
GRB\,971214       & $120\pm 275$ & \nodata 	& 3 \\
GRB\,980329$^{a}$ & $460\pm 210$ & $615\pm 275$ & \nodata \\
GRB\,980613  	  & $380\pm 200$ & $50\pm 140$ 	& \nodata \\
GRB\,980703  	  & $<380$ 	 & $180\pm 25$ 	& 10 \\
GRB\,981226$^{b}$ & \nodata	 & $150\pm 85$ 	& \nodata \\
GRB\,990510  	  & \nodata	 & $190\pm 750$ & \nodata \\
GRB\,990705  	  & \nodata	 & $190\pm 165$ & \nodata \\
GRB\,991208  	  & $<370$ 	 & $70\pm 30$ 	& 20 \\
GRB\,991216  	  & $<395$ 	 & $80\pm 70$ 	& \nodata \\
GRB\,000131  	  & \nodata	 & $9800\pm 6070$ & \nodata \\
GRB\,000210  	  & $560\pm 165$ & $90\pm 45$ 	& 3 \\
GRB\,000301C 	  & $<670$ 	 & $640\pm 270$ & \nodata \\
GRB\,000418  	  & $690\pm 195$ & $330\pm 75$ 	& 55 \\
GRB\,000911  	  & $495\pm 195$ & $85\pm 70$ 	& 2 \\	 	 
GRB\,000926  	  & \nodata	 & $820\pm 340$ & \nodata \\
GRB\,010222  	  & $610\pm 100$ & $300\pm 115$ & 1.5 \\
GRB\,011211  	  & $350\pm 255$ & \nodata	& \nodata
\enddata
\tablecomments{The columns are (left to right), (1) Source name, (2)
SFR derived from the sub-mm flux, and (3) SFR derived from the radio
flux, and (4) SFR derived from optical observations.  The upper limits
represent $2\sigma$ values in the case
when the measured flux was negative (see Table~\ref{tab:submm}).}
\end{deluxetable}

\clearpage
\begin{figure} 
\epsscale{1}
\plotone{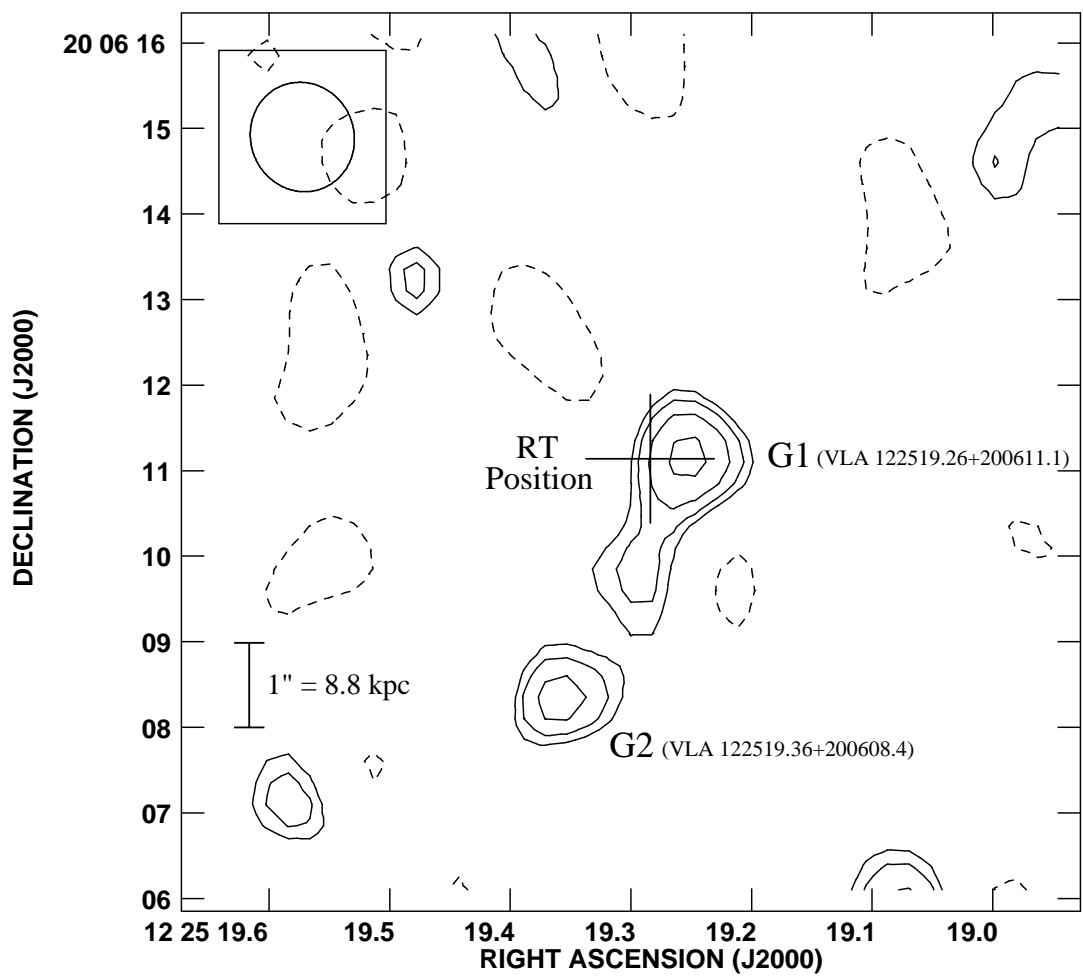}
\caption{Contour plot of a $5\times 5$ arcsec$^2$ field observed at
1.43 GHz and centered on the position (Berger et al. 2001) of the
radio transient associated with GRB\,000418 (marked by cross).
Contours are plooted at $-2^{1/2}, 2^{1/2}, 2^1, 2^{3/2}, 2^2, \,{\rm
and}\,2^{3/2}$.  Source G1 is the host galaxy of GRB\,000418, while
source G2 is a possible companion galaxy.  In addition, there appears
to be a bridge of radio emission connecting galaxies G1 and G2.  A
comparison to the synthesized beam (upper left corner) reveals that G1
and G2 are slightly extended.
\label{fig:lmap}}
\end{figure}

\clearpage
\begin{figure} 
\epsscale{0.7}
\plotone{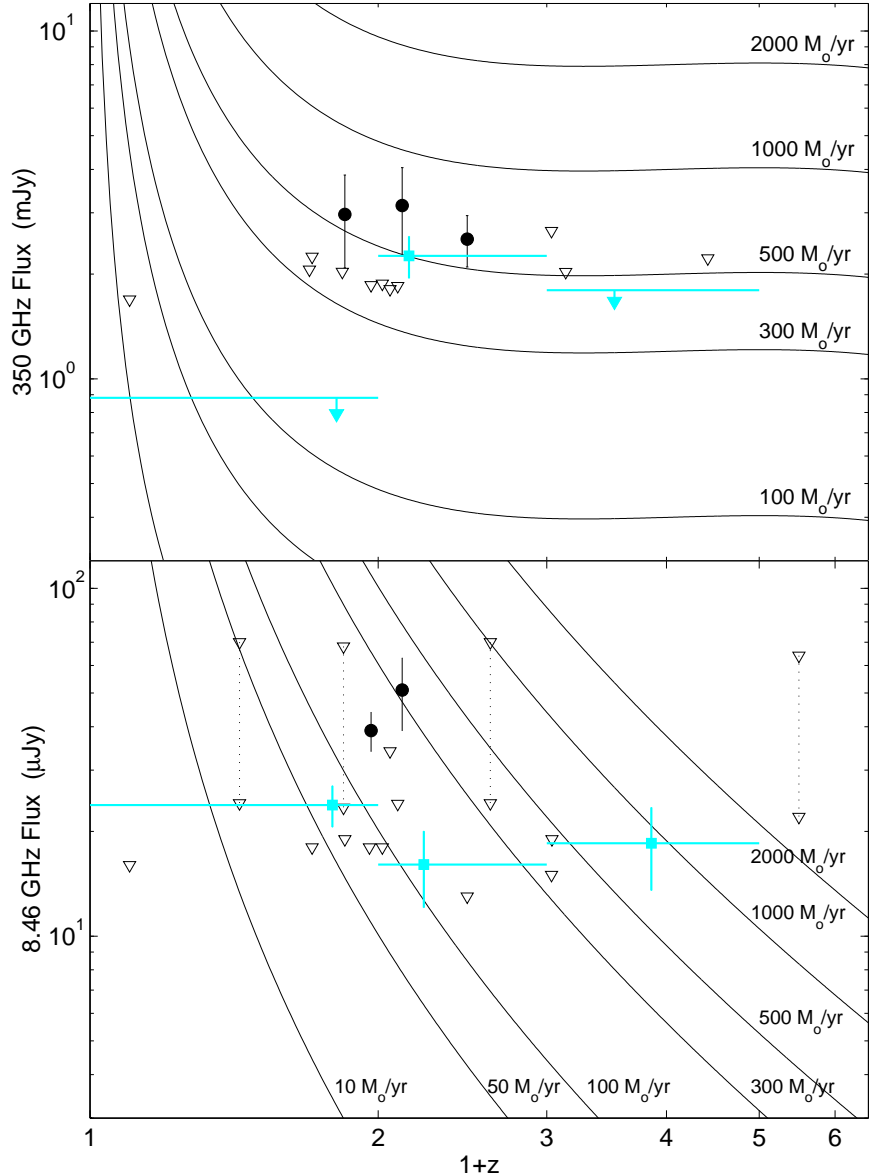}
\caption{Sub-mm (top) and radio (bottom) fluxes for 20 GRB host
galaxies plotted as a function of source redshift.  The solid symbols
are detections ($S/N\!>\!2$ in the sub-mm
and $S/N\!>\!3$ in the radio), while the inverted triangles are
$2\sigma$ upper limits.  In the bottom panel, the upper limits linked
by dotted lines are the upper limits from the ATCA observations at 1.4
GHz (upper triangles) converted to 8.46 GHz (lower triangles) using
$F_\nu\propto \nu^{-0.6}$.  Also plotted are the ATCA upper limit for
GRB\,990712 ($z=0.433$; Vreeswijk et al. 2002), the VLA detection of
the host of GRB\,980703 (Berger et al. 2001), and the sub-mm detection
of the host of GRB\,010222 (Frail et al. 2001).  The source at
$1+z=1.2$ in both panels is the host of GRB\,980329 which does not
have a measured redshift.  The points and upper limits with horizontal
error bars are weighted average fluxes in the redshift bins:
$0\!<\!z\!<\!1$, $1\!<\!z\!<\!2$, and $z\!>\!2$.  Finally, the thin
lines are contours of constant star formation rate (using 
Equation~\ref{eqn:sfr} with the parameters specified in \S\ref{sec:sfr}). 
\label{fig:fluxes}}
\end{figure}

\clearpage
\begin{figure}
\epsscale{1} 
\plotone{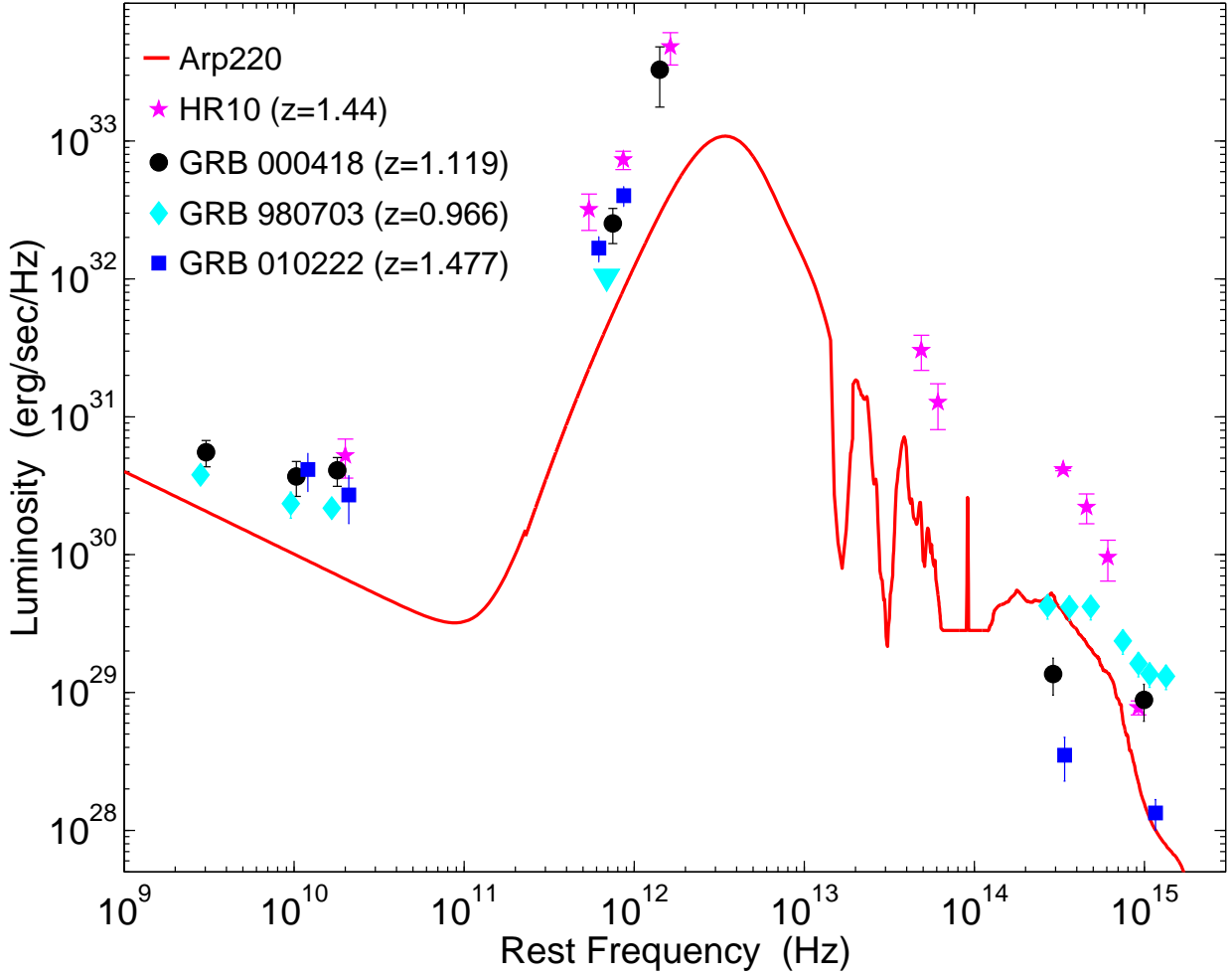}
\caption{SEDs of the host galaxies of \grb{}, GRB\,980703, and
GRB\,010222 compared to the SED of the local starburst galaxy
Arp\,220, and the high-$z$ starburst galaxy HuR\,10.  The luminosities
are plotted at the rest frequencies to facilitate a direct comparison.
The GRB host galaxies are more luminous than Arp\,220, and are similar
to HuR\,10, indicating that their bolometric luminosities exceed
$10^{12}$ L$_\odot$, and their star formation rates are of the order
of $500$ M$_\odot$ yr$^{-1}$.  On the other hand, the spectral slopes
in the optical regime are flatter than both Arp\,220 and HuR\,10,
indicating that the GRB host galaxies are bluer than Arp\,220 and
HuR\,10.  
\label{fig:sed}}
\end{figure}

\clearpage
\begin{figure}
\epsscale{1} 
\plotone{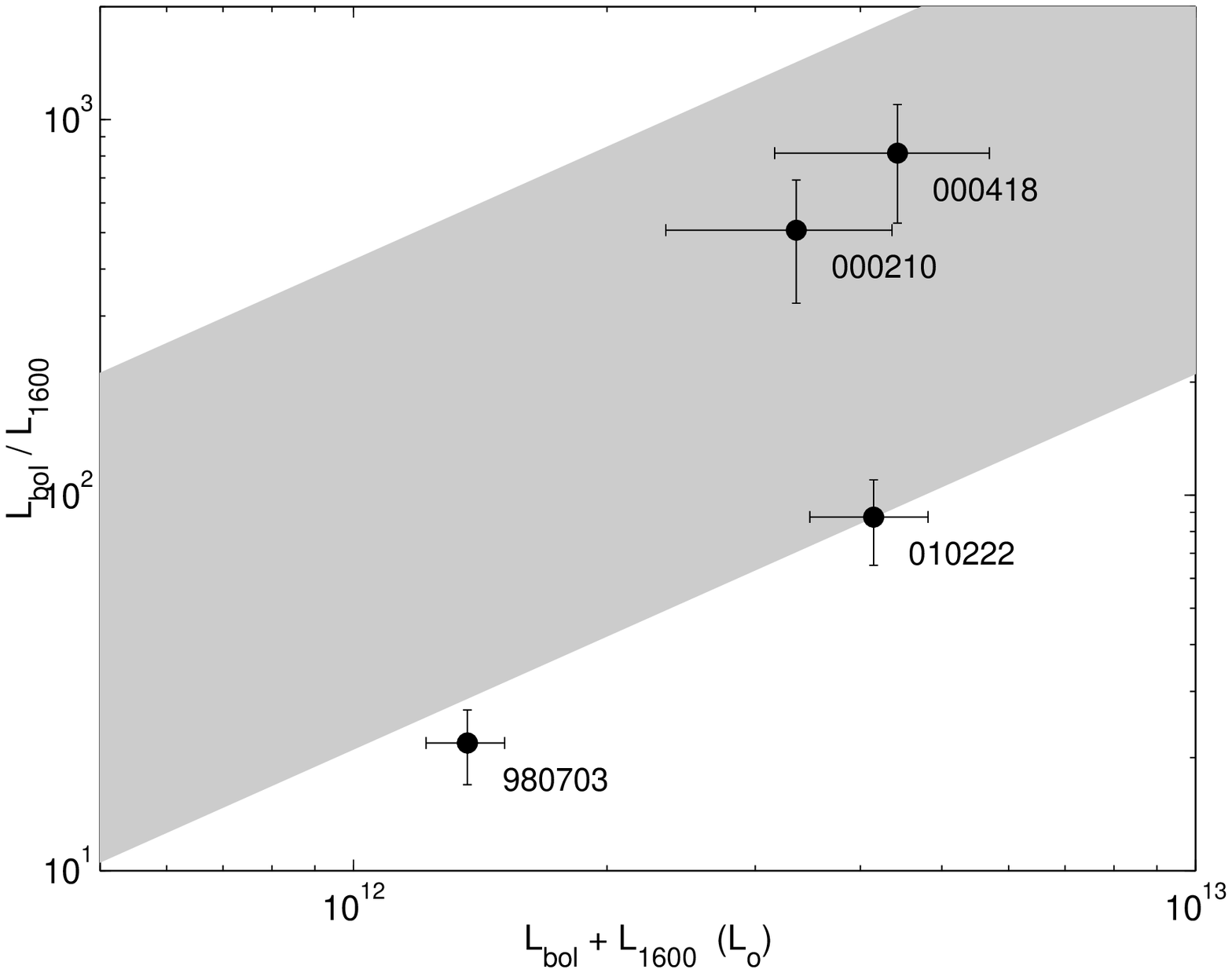}
\caption{Ratio of bolometric luminosity, $L_{\rm bol}$ to luminosity
at $1600$\,\AA{}, $L_{1600}$ plotted as a function of the combined
luminosity.  The ordinate provides a measure of the amount of dust
obscuration, while the abscissa provides a measure of the total star
formation rate.  Black circles are the host galaxies detected here and
by \citet{bkf01} and \citet{fbm+02}, while the shaded region is from
\citet{as00} based on observations of starbursts and ULIRGs at $z\sim
1$.  Clearly, there is a trend in both cases for more dust obscuration
at higher star formation rates, but the level of obscuration in GRB
hosts is significantly lower than typical starbursts at the same
redshift. 
\label{fig:lbol}}
\end{figure}

\clearpage
\begin{figure}
\epsscale{1} 
\plotone{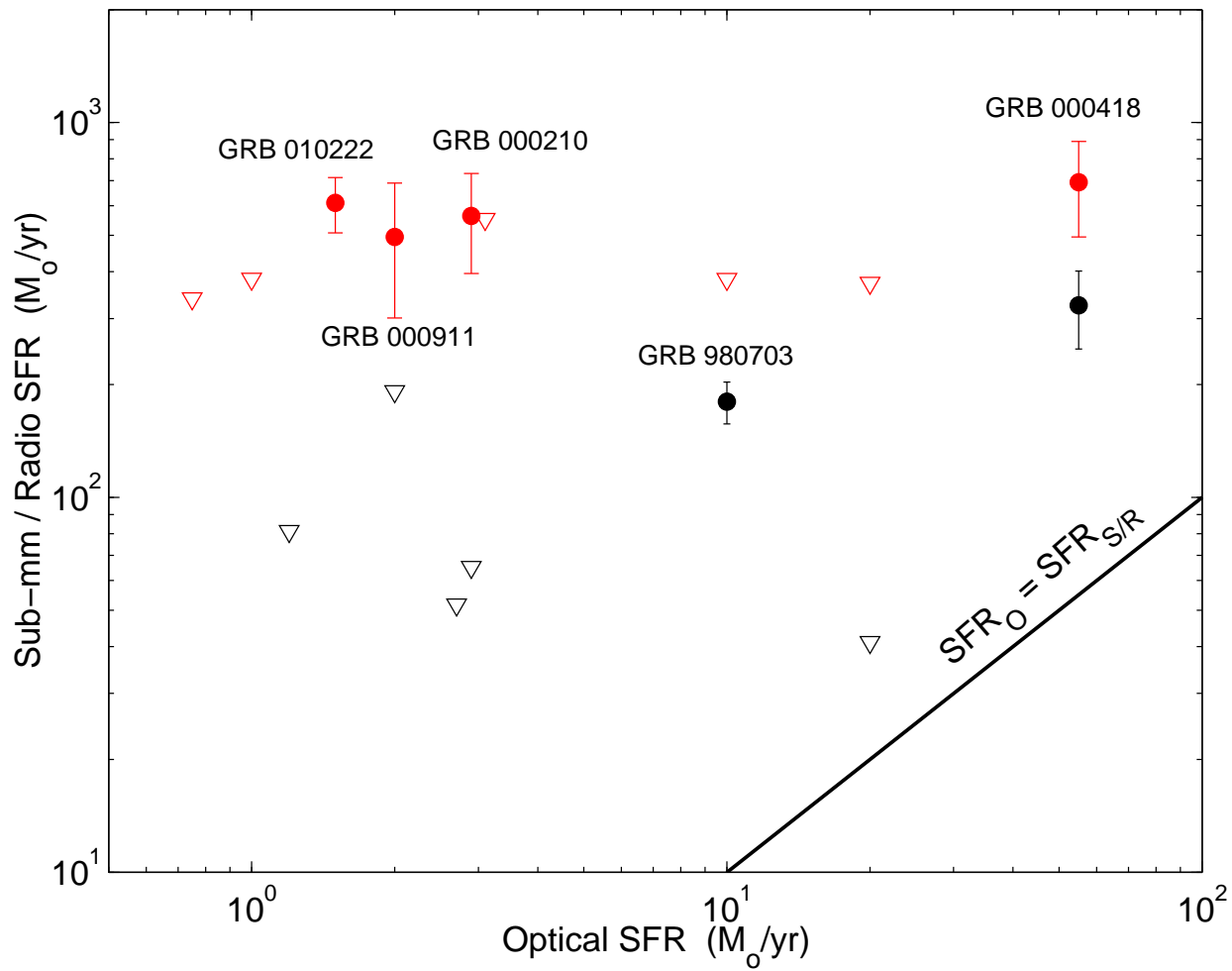}
\caption{Sub-mm/radio vs.~optical star formation rates for several GRB
host galaxies.  The line in the bottom right corner designates a
one-to-one correspondence between the two SFRs.  Clearly, the hosts
that gave appreciable sub-mm and/or radio flux have a large fraction
of obscured star formation.  In fact, the GRB hosts with a higher dust
bolometric luminosity have a higher fraction of obscured star
formation. 
\label{fig:optsubmm}}
\end{figure}

\clearpage
\begin{figure}
\epsscale{1} 
\plotone{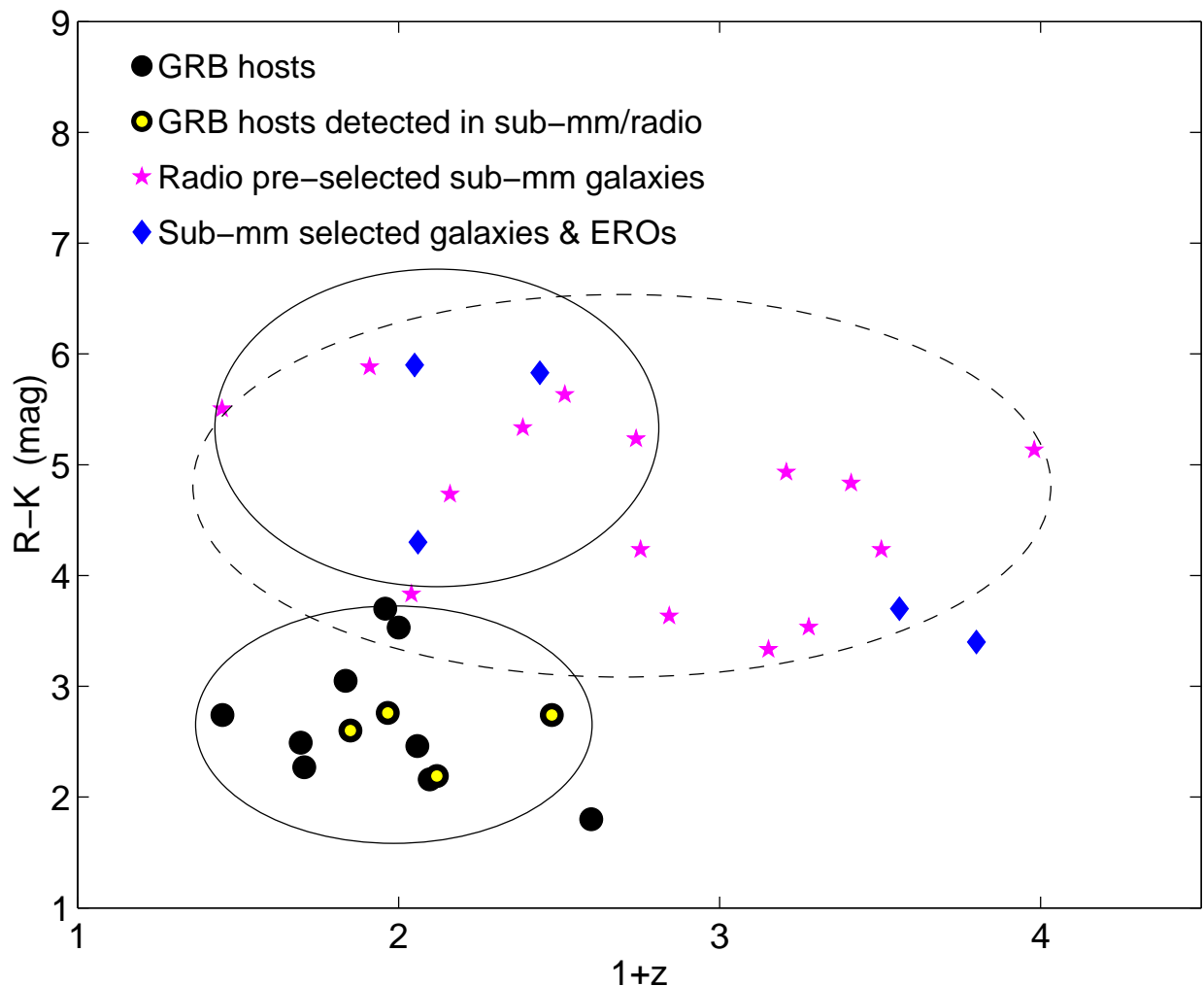}
\caption{$R-K$ color as a function of redshift for GRB host galaxies,
and radio pre-selected sub-mm selected (Chapman et al. 2002).  The
solid ellipses are centered on the mean color and redshift for each
population of galaxies in the redshift range $z<1.6$, and have widths
of $2\sigma$.  The dashed ellipse is the same for the sub-mm
population as a whole.  Clearly, the GRB hosts are significantly bluer
than the sub-mm galaxies in the same redshift range, indicating a
possible preference for younger star formation episodes in GRB
selected galaxies.
\label{fig:color}}
\end{figure}

\clearpage
\begin{figure} 
\epsscale{0.8}
\plotone{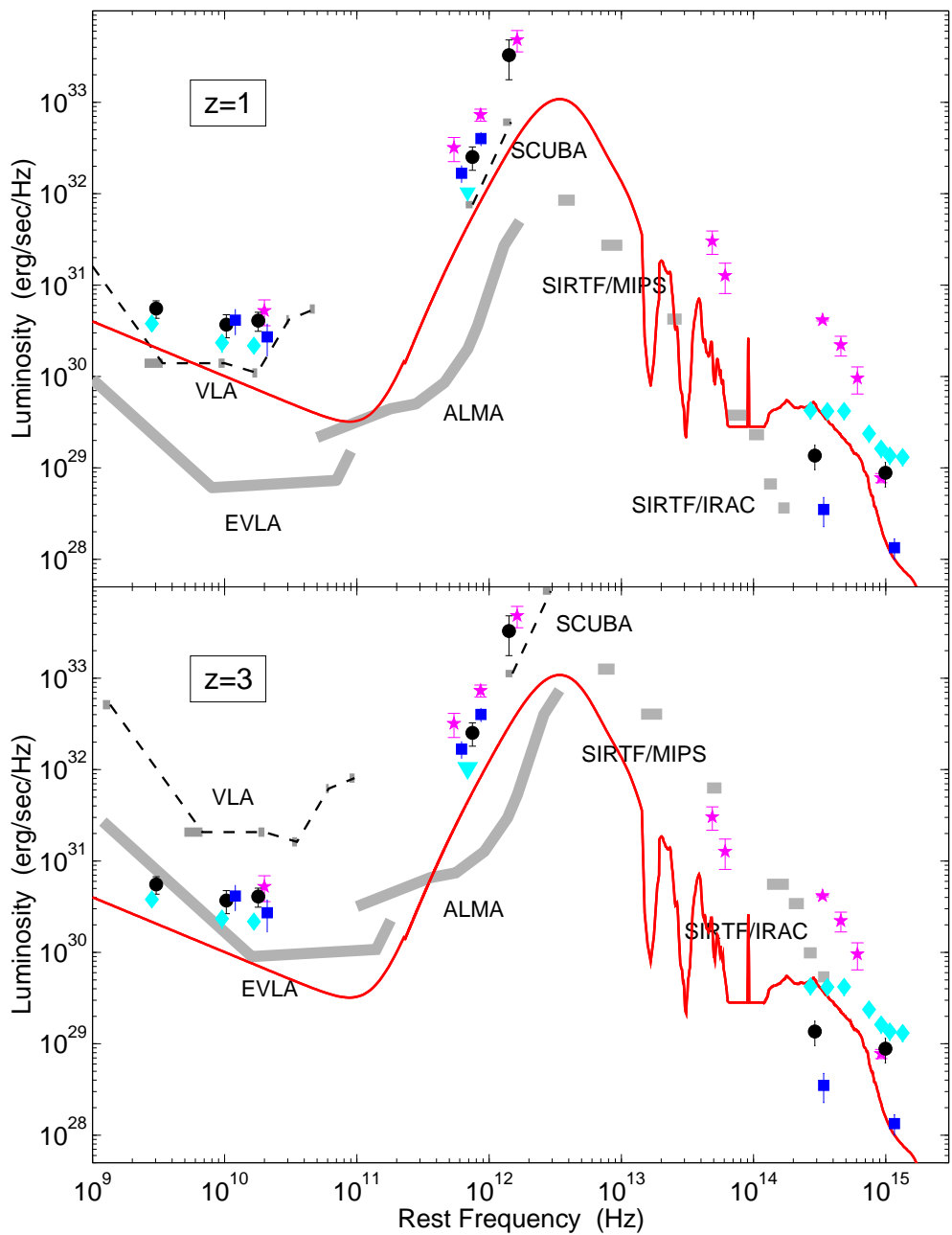}
\caption{Same as Figure~\ref{fig:sed}, overplotted with the EVLA, ALMA,
and SIRTF bands at $z=1$ and $z=3$.  The shaded regions correspond to
the $1\sigma$ sensitivity in a 200 sec exposure for each instrument,
while the dashed lines are the typical $1\sigma$ sensitivities for
current instruments (i.e.~VLA and SCUBA).  Clearly, the new
observatories will allow a significant increase in sensitivity, and
spectral coverage over current instruments.  As a result, the
radio/sub-mm/IR observations will be able to probe lower luminosity
(and hence more typical) star-forming galaxies. 
\label{fig:sirtf}}
\end{figure}

\end{document}